\RequirePackage{fix-cm}
\documentclass[twocolumn]{svjour3}          
\smartqed  
\usepackage{psfrag}
\usepackage{amsmath}
\usepackage{url,verbatim}
\usepackage{amsmath,psfrag}
\usepackage{amsfonts}
\usepackage{graphicx}
\usepackage{framed}
\usepackage{xcolor}
\usepackage{tikz}
\usepackage{caption}
\usepackage{array}
\begin{document}

\title{Synchronisation in TCP networks with Drop-Tail Queues}


\author{Nizar~Malangadan\and Gaurav~Raina\and Debayani~Ghosh 
}

\institute{N.~Malangadan \and G.~Raina \at
              Department of Electrical Engineering\\
              Indian Institute of Technology Madras, Chennai             \\ \email\{ee11s040, gaurav\}@ee.iitm.ac.in           
           \and
           D.~Ghosh \at
              Electronics and Communication Engineering Department\\ Thapar Institute of Engineering and Technology, Patiala\\
              \email{debayani.ghosh@thapar.edu}
}

\date{Received: date / Accepted: date}

\maketitle

\begin{abstract}
The design of transport protocols, embedded in end-systems, and the choice of buffer sizing strategies, within network routers, play an important role in performance analysis of the Internet. In this paper, we take a dynamical systems perspective on the interplay between fluid models for transport protocols and some router buffer sizing regimes. Among the flavours of TCP, we analyse Compound, as well as Reno and Illinois. The models for these TCP variants are coupled with a Drop-Tail policy, currently deployed in routers, in two limiting regimes: a small and an intermediate buffer regime. The topology we consider has two sets of long-lived TCP flows, each passing through separate edge routers, which merge at a common core router. Our analysis is inspired by time delayed coupled oscillators, where we obtain analytical conditions under which the sets of TCP flows synchronise. These conditions are made explicit in terms of coupling strengths, which depend on protocol parameters, and on network parameters like feedback delay, link capacity and buffer sizes. We find that variations in the coupling strengths can lead to limit cycles in the queue size. Packet-level simulations corroborate the analytical insights. For design, small Drop-Tail buffers are preferable over intermediate buffers as they can ensure both low latency and stable queues.
\subclass{MSC code1 \and MSC code2 \and more}
\end{abstract}

\section{Introduction}
\label{intro}
The Internet is perhaps the largest manmade engineered system, as viewed from the perspective of the number of users that it supports, and the number of routers needed to forward and route information. Developing mathematical models to develop an understanding of the underlying dynamics, and performance, such a network is a challenging task. The network could be viewed as a large queuing network as also a large feedback system, where explicit or implicit feedback from the network helps to regulate the way transport protocols send traffic into the network.   

In a broader context, understanding the impact of delayed information on the dynamics of queuing systems is beginning to generate significant research interest in the literature \cite{corella,pender4,pender2,pender3,pender11}. Some common scenarios where delayed information can potentially hurt customer experience are the Internet, mobile services, healthcare services and public transportation, where it is important to communicate the waiting times or the queue length information with the customers \cite{Doldo,pender5,pender6}. In this paper, we study queuing systems which arise in the Internet, wherein majority of the traffic is regulated by Transmission Control Protocol (TCP), and congestion notifications are delayed. Our modelling perspective is guided by a dynamical systems approach, where appropriate fluid models for limiting queuing regimes interplay with delayed feedback.  

Queuing delays can build quite easily due to large and unmanaged router buffers~\cite{bufferbloat}. Such delays are reportedly on the rise, and they can hurt network performance and Quality of Service (QoS)~\cite{queue_delay}. If one were to control such delays, it would be important to study the interaction of transport protocols along with queue management policies. In this paper, our focus is on the analysis of three flavours of the Transmission Control Protocol (TCP): Compound TCP~\cite{compound}, which is the default protocol in the Windows operating system, TCP Reno~\cite{renoTowsley,rtw}, a standardised protocol, and TCP Illinois~\cite{illinois}, a new proposal. TCP requires feedback, from the network, to efficiently perform flow and congestion control. So a model should also consider methods for queue management in network routers. In our work, the models for these TCP variants are coupled with a Drop-Tail queue management policy, currently deployed in routers, in two different limiting regimes: a \emph{small buffer} and an \emph{intermediate buffer} regime.         

For some flavours of TCP, it is possible to devise fluid models which inturn can help guide the choice of system parameters. These models are usually deterministic, non-linear, and time delayed equations, whose stability and bifurcation properties can guide network performance studies. For such \textit{fluid models} of Compound TCP see~\cite{ccdc,raja}, for TCP Reno see~\cite{hhtc,rtw,rw}, and for TCP Illinois see~\cite{raja}. Queue management, in routers, has been an active area of research in the networking community but the simple Drop-Tail queue policy continues to be widely used in practice. Drop-Tail simply drops all incoming packets after the router buffer is full. For the development of some Drop-Tail models, with different buffer sizing regimes, see~\cite{rw}. Quite naturally, models of TCP should be studied in conjunction with models of queue management to develop a system level understanding. 

A large component of queuing delays, in the network, are being caused by \emph{large} router buffers that do not employ effective queue management strategies. To that end, we consider models for intermediate and small buffers. Reducing the size of buffers will reduce latency, but without clear design guidelines we may witness the emergence of oscillations and synchronisation effects in the queue size dynamics, which in turn could lead to the loss of link utilisation. A key focus in this paper is to explore and understand this phenomena of synchronisation. In a different context, the authors in \cite{pender2,pender11} study a class of fluid models of queues with delayed feedback and conclude that oscillations emerge if the delay gets too large.

Over the years, a large body of research has been motivated by work on the synchronisation of oscillators; for example, see~\cite{kuramoto1,kuramoto2,tousi,woods}.
An intriguing phenomena in such systems is the spontaneous locking, by the oscillators, into a common frequency.
Our style of analysis is inspired from the study of the time delayed Kuramoto model for coupled oscillators~\cite{sync2,strogatz}. The model was originally developed for coupled \emph{biological oscillators}, such as flashing fireflies \cite{buck} and cardiac pacemaker cells \cite{peskin}. The model continues to provide inspiration in different application areas. For example, recent work on power systems has also been influenced by the Kuramoto model~\cite{syncpower1,syncpower2}. This style of analysis has been used to study the synchronisation of TCP Reno flows coupled with small Drop-Tail buffers~\cite{hhtc}. In our previous work~\cite{ccdc}, we studied Compound TCP coupled with a Drop-Tail queue policy in a small and an intermediate buffer regime. We now generalise the analysis which is readily applicable to different fluid models of TCP. In particular, we also study TCP Reno which is a standardised version of TCP and TCP Illinois which is a new TCP proposal. Compared to~\cite{hhtc}, our work generalises the models and analysis to three flavours of TCP combined with both small and intermediate Drop-Tail buffers. Even though Compound TCP is widely deployed in the Internet, very few analytical models exist for Compound. A key insight obtained is the impact of protocol parameters and the buffer size on the analytical conditions for synchronisation. To that end, we highlight the role of the parameters embedded in Compound. Additionally, using packet-level simulations, we exhibit the emergence of synchronised phenomena, in the form of limit cycles, which corroborates the analytical insights. Limit cycles, in the queue size dynamics, were not exhibited in~\cite{hhtc}. 

Our model setting is as follows. We consider two distinct sets of long lived TCP flows, each regulated by separate edge routers, which merge at a common core network router.
For both the small and the intermediate buffer regimes, we derive analytical conditions under which \emph{synchronisation} would occur. The conditions are made explicit in terms of a coupling strength, which depend on TCP parameters and on network parameters like the feedback delay, link capacity and buffer sizes. It is shown that variations in the coupling strength can lead to non-linear oscillations in the form of limit cycles in the queue size. Such synchronisation phenomena in network queues would affect end-to-end network performance, and should be avoided. Packet-level simulations, conducted using the Network Simulator (NS2)~\cite{ns2}, corroborate our analytical insights. Small buffers, with Drop-Tail, have low latency and can be designed to ensure that the queue size dynamics do not exhibit limit cycles. Thus small router buffers are an appealing design option for future network architectures.    

The rest of this paper is structured as follows. In Section \ref{section:TCPs}, we briefly describe the algorithms of the various TCP flavours. In Section \ref{section:TCPFluidModel}, fluid models for the dynamics of the TCP variants with Drop-Tail are outlined. We derive expressions for coupling strengths in Section \ref{section:coupled}, and analytical conditions for synchronisation in Section \ref{section:synchronisation}. In Section \ref{section:simulations} we conduct some packet-level simulations, and summarise our contributions in Section \ref{section:conclusion}.

\newcommand{\mainw}{\mathsf{w}}
\newcommand{\w}[1]{\mainw_{#1}}
\newcommand{\weqC}[1]{\widetilde{\mainw}_{#1}^*}
\newcommand{\dw}[1]{\varDelta\mainw_{#1}}
\newcommand{\weq}[1]{\mainw^*_{#1}}
\newcommand{\myi}{j}
\newcommand{\myj}{i}
\newcommand{\mym}{m}
\newcommand{\myf}{g}
\section{Transport protocols}
We now briefly outline the congestion control algorithms for the following transport protocols: Compound TCP~\cite{compound}, TCP Reno~\cite{renoTowsley,rtw} and TCP Illinois~\cite{illinois}. 
\label{section:TCPs}
\subsection{Compound TCP}
\label{subsection:CTCPalgorithm}
Compound is the default transport layer protocol  of the Windows operating system, which was introduced as part of the Windows Vista and Windows
Server 2008 TCP stack~\cite{compound}.
Compound TCP uses both \emph{queuing delay} and \emph{packet loss} for managing its flow control algorithms.
To describe the key algorithms of Compound, we use the same notation which was used in the original paper~\cite{compound}.
In Compound, the sending window $\w{}$, which regulates the flow rate, is calculated as follows:
\[\w{} = \min\bigl(cwnd+dwnd, awnd\big), \]
where $cwnd$ is the loss window, $dwnd$ is the delay window, and $awnd$ is the minimum window size. 
To estimate the transmission delay of packets, a state variable called $baseRTT$ \big(base round-trip time (RTT)\big) is maintained. 
The number of backlogged packets of the connection ($diff$) can be estimated by the following 
\begin{equation*}
 diff = \bigg( \frac {\w{}}{baseRTT} - \frac {\w{}}{RTT} \bigg) baseRTT.
\end{equation*}
The algorithm to determine $dwnd$ is as follows
%
\begin{small}
\begin{displaymath} 
dwnd(t+1) =  \begin{cases}
dwnd(t)+\Big(\alpha\,\w{}(t)^k-1\Big)^+ & diff < \gamma\\
\bigl(dwnd(t)-\zeta\,diff\bigr)^+  & diff \ge \gamma\\
\Big(\w{}(t)(1-\beta)-cwnd/2\Big)^+ & \text{if loss},
\end{cases}  
\end{displaymath}
\end{small}
where $(z)^+$ is defined as $\max(z,0)$ and $\zeta > 0$. 
The parameters  $\alpha$, $\beta$ and $k$, influence the scalability, smoothness and responsiveness of the window function respectively,  
and $\gamma$ is a protocol threshold. 
The default parameter values are $\alpha = 0.125$, $\beta = 0.5$, $k = 0.75$, $\gamma=30$ and $\zeta=0.5$~\cite{compound}. 
If $diff<\gamma$, the network path is assumed to be underutilised and Compound acts aggressively by increasing the sending rate. If $diff\geq\gamma$, 
the delay based
component, $dwnd$, is gracefully reduced. In the event that a packet loss is detected, Compound TCP reduces its congestion window by half, similar to Reno.

\subsection{TCP Reno}
TCP Reno is a standard implementation of the Transmission Control Protocol. 
The protocol relies only on packet loss as the 
feedback signal to regulate its flow and congestion control mechanisms.
 The response for the feedback is defined by  the Additive Increase Multiplicative Decrease (AIMD) algorithm, 
where the window size $\w{}$ is updated as follows
\begin{small}
\begin{displaymath} 
\w{}(t_{n+1}) =  \begin{cases}
\w{}(t_n)+ \dfrac{1}{\w{}(t_n)} &\!\!\! \text{if positive acknowledgement}\\[1.5ex]
\dfrac{\w{}(t_n)}{2} &\!\!\! \text{if loss},
\end{cases}  
\end{displaymath}
\end{small}
%
where the sequence $t_n$  corresponds to the time instants at which either successful acknowledgements, or loss of packets are detected.

%
\subsection{TCP Illinois}

TCP Illinois~\cite{illinois}, like Compound TCP, uses both queuing delay and packet loss for flow and congestion control. 
The protocol aims to achieve concave-shaped curves for the window by varying the parameters 
$\alpha$ and $\beta$ of the standard Additive Increase Multiplicative Decrease (AIMD) TCP Reno  algorithm, based on the observed average queuing delay, $d_a$. These protocol parameters vary according to $\alpha = f_1(d_a)$ and $\beta=f_2(d_a )$,
where $f_1(d_a)$ is an increasing function and $f_2(d_a)$ is a decreasing function. The reader is referred to~\cite{illinois} for a detailed description of these functions.

The additive increase and multiplicative decrease
parameters of TCP Illinois also depend on the delay, but the regime we focus on is where the the queuing delay forms a negligible component of the end to end delay.
In this regime, Illinois uses $\alpha=\alpha_{max}$
and $\beta=\beta_{min}$, which would result in the  following algorithm:

\begin{small}
\begin{displaymath} 
\w{}(t_{n+1}) =  \begin{cases}
\w{}(t_n)+ \dfrac{\alpha_{max}}{\w{}(t_n)} &\!\!\! \text{if positive acknowledgement}\\[1.5ex]
(1-\beta_{min})\,\w{}(t_n) &\!\!\! \text{if loss}.
\end{cases}  
\end{displaymath}
\end{small}
The default values of the parameters are $\alpha_{max}=10$ and $\beta_{min}=0.125$~\cite{illinois}.

\section{Fluid models}
\label{section:TCPFluidModel}
We now outline fluid models for the transport protocols in small and intermediate buffer sizing regimes. 
\subsection{Generalised TCP}
\label{subsection:FluidmodelGeneralisedTCP}
Consider two distinct sets of TCP flows that are regulated by separate edge routers with Drop-Tail buffers. 
For the $\mym^{th}$ set of flows ($\mym=1,2$), let the \textit{average window size} be $\w{\mym}(t)$ and the round-trip time be $\tau_\mym$. Then the 
\textit{average rate} of flows $x_\mym(t)=\w{\mym}(t)/\tau_\mym$.
Further, consider that  TCP increments the window size $\w{}$ of a flow by $i(\w{})$ per  positive acknowledgement, and decrements it by
$d(\w{})$ when a packet drop is detected. 
Then from~\cite{rw},  a \emph{many flows} fluid model for a generalised TCP is 
%

\begin{align}
 \dot{\w{}}_\mym(t) =&\, \frac{\w{\mym}(t-\tau_\mym)}{\tau_\mym}\biggl( i\bigl(\w{\mym}(t)\bigr)\bigl(1-p_\mym(t-\tau_\mym)\bigr)\label{eq:tcpeq} \\ -& d\bigl(\w{\mym}(t)\big)p_\mym(t-\tau_\mym)\biggr)\,\, ;\,\, \mym=1,2,\notag
\end{align}

where 
$p_\mym(.)$ is the edge buffer
packet loss probability
for the $\mym^{th}$ set of TCP flows.
Whenever there is a packet loss, which would occur with a probability of $p_m(.)$, the window size of a TCP flow decrements. The window size increases when there is no packet loss, which would occur with a probability of $\big(1-p_m(.)\big)$. Thus, multiplying $i(\w{})$ by $\big(1-p_m(.)\big)$ makes the model more realistic in capturing the evolution of the window size dynamics. This would also impact the non-linear structure of the fluid model. This aspect was not incorporated in previous models \cite{hhtc,ccdc,raja}.
Throughout this paper, $\dot{z}(t)$ denotes the derivative of $z(t)$ with respect to time $t$.
The functional forms for $i(\w{})$ and $d(\w{})$ depend on the congestion avoidance algorithms of the transport protocols.
The functional forms of $i(\w{})$ and $d(\w{})$ for TCP Reno are derived in~\cite{rw}, and for Compound and Illinois in~\cite{raja}. 
The functional forms for these protocols are listed in Table~\ref{table:windowfunctions}. Substituting these functional forms in \eqref{eq:tcpeq} will give a fluid model for the corresponding
transport protocols. We now outline the models for the packet loss probability at the bottleneck queue, in small and intermediate buffer regime. For the traffic, we consider both smooth and bursty traffic.
\\
\begin{table}[t]
\centering
 \begin{tabular}{lccc}
  \hline\\[-1ex]
  TCP &  $ i(\w{})$ &  $d(\w{})$      \\[1ex]
  \hline
  \hline\\[-0.5ex]
   \textit{Compound} & $\displaystyle\,\,\,\,\, \alpha \w{}^{k-1}$ & $\displaystyle \beta \w{}$  &\\[3ex]
 
   \textit{Reno} &  $\displaystyle \frac{1}{\w{}}$ & $\displaystyle \frac{\w{}}{2}$    \\[3ex]
    \textit{Illinois} &  $\displaystyle \frac{\alpha_{max}}{\w{}}$ & $\displaystyle \beta_{min}\w{}$   \\[2ex]

  \hline
  \hline
 \end{tabular}
 \vspace{8pt}
\caption{ Window increase and decrease functions of the transport protocols. The default values of the protocol parameters are: $\alpha=0.125$, $\beta=0.5$ and $k=0.75$ (for Compound); 
$\alpha_{max}=10$ and $\beta_{min}=0.125$ (for Illinois). }
\label{table:windowfunctions}
\end{table}

\noindent\textit{Small buffers and Drop-Tail}\\

\noindent\textit{Smooth traffic:} In the smooth traffic scenario, the arrival process at the bottleneck queue constitutes a point process, wherein each point represents one packet. When the access links are much slower as compared to the core link, the packets are naturally spaced out, thus ensuring smooth traffic flow into the bottleneck queue. In real networks, the core routers typically operate at a much faster rate than the edge routers and hence the incoming traffic to the core router is inherently smooth \cite{S2,Lakshmikantha}. In such a scenario, with a large number of TCP flows, the blocking probability of an M/M/1
queue has been proposed as a simple fluid model for small Drop-Tail buffers \cite{Poisson_limit,rw}. Thus, with a large number of flows, the edge routers may be represented by the following model:  
\begin{equation}
 p_\mym\big(\w{\mym}(t)\big) = \left(  \frac{\w{\mym}(t)}{c'_\mym\tau_\mym} \right)^{b_\mym}, 
\label{eq:edgeSmall}
\end{equation}
where $b_\mym$ is the router buffer size, $c'_\mym$ is the link capacity per flow, $\tau_\mym$ is the round-trip time and $x_\mym(t)$ is the rate, where 
$\w{\mym}(t) =x_\mym(t)\tau_\mym$. \\

\noindent\textit{Bursty traffic:} In the bursty traffic scenario, the arrival process at the bottleneck queue can be thought of as a point process. However, each point now represents a batch arrival, \emph{i.e.}, each point represents a clump of packets. This traffic scenario can be realised by operating the access links at faster speeds \cite{Ayesta,S2,Lakshmikantha}. In \cite{Wischik}, the author has modified the blocking probability of an M/M/1/B queue to propose a crude approximation for the packet loss probability in presence of bursty traffic. Note that the model for the packet drop probability is derived by assuming that if the first packet of a batch gets dropped, then all successive packets get dropped.This yields the following fluid model for the edge routers \cite{Wischik}: 
\begin{equation}
 p_\mym\big(\w{\mym}(t)\big) =\frac{1}{\w{\mym}(t)} \left(  \frac{\w{\mym}(t)}{c'_\mym\tau_\mym} \right)^{\frac{b_\mym}{\w{\mym}}}, 
\label{eq:edgeSmall_bursty}
\end{equation}
Here, $\w{\mym}$ also represents the batch size at time instant $t.$ To simply analysis, in our work, we assume that the batch size is a constant, and refer to it as the \emph{burstiness parameter}. Let the burstiness parameter corresponding to the edge routers be $n_{\mym};$ $m=1,2.$ Then, the packet loss probabilities at the edge routers can be re-written as follows:
\begin{equation}
 p_\mym\big(\w{\mym}(t)\big) =\frac{1}{n_{\mym}} \left(  \frac{\w{\mym}(t)}{c'_\mym\tau_\mym} \right)^{\frac{b_\mym}{n_\mym}}, 
\label{eq:edgeSmall_burstyconstant}
\end{equation}
Note that when the burstiness parameter $n_\mym=1,$ it yields the fluid models with smooth traffic. Thus, the model \eqref{eq:edgeSmall_burstyconstant} can be thought of as a generalisation of \eqref{eq:edgeSmall}. Hence, we analyse \eqref{eq:tcpeq} coupled with the packet loss probability model \eqref{eq:edgeSmall_burstyconstant}. Assume $\w{\mym}^*$ to be the equilibrium of the system \eqref{eq:tcpeq} and \eqref{eq:edgeSmall_burstyconstant}.
Then, at equilibrium, we have
%
%
\begin{align*}
 i(\weq{\mym})\bigl(1-p_\mym(\weq{\mym})\bigr) = d(\weq{\mym})p_\mym(\weq{\mym}).
\end{align*}
Let $\w{\mym}(t)=\w{\mym}^*+\varDelta{}\w{\mym}(t)$, and linearising the non-linear equations about the equilibrium, we get the following system
\begin{equation}\small
          \varDelta{}\dot{\w{}}_\mym(t) =  -\frac{i(\weq{\mym})}{\tau_\mym}\biggl( g\big(\weq{\mym}\big)\dw{\mym}(t)+ \frac{b_\mym}{n_\mym}\dw{\mym}(t-\tau_\mym) \biggr), \label{eq:GenTCPLinearSmallBuffer}
\end{equation} 
%
where
{\small\begin{equation*}
 \myf(\weq{\mym})=\left(\frac{\weq{\mym}d'(\weq{\mym})}{d(\weq{\mym})} - \frac{\weq{\mym}i'(\weq{\mym})}{i(\weq{\mym})}\right)\frac{d(\weq{\mym})}{i(\weq{\mym})+d(\weq{\mym})}. 
\end{equation*}}
\hspace{-3pt}We now derive the intrinsic frequency of the model.\\


%
\noindent\textit{Intrinsic frequencies:} 
In the multi-bottleneck scenario, it has been shown that with Compound TCP, the dynamical system would undergo a Hopf bifurcation as the underlying fluid model transits from a locally stable to a locally unstable state~\cite{debayani1,debayani2}. The existence of a Hopf highlights that one would expect to observe limit cycles in the neighbourhood of the locally unstable regime~\cite{nonlinear}. While it has not exhibited that with TCP Reno and Illinois, a Hopf bifurcation would occur, we assume that we would also expect non-linear oscillations in the form of limit cycles as the system loses local stability. Packet-level simulations, conducted later in the paper, do confirm the existence of such limit cycles. 
We approximate these emerging limit cycles with harmonic oscillations. 
To obtain their frequency, we substitute $\varDelta{}\w{\mym}(t) = r_\mym e^{\myi\overline{\theta}_\mym(t)}$ into (\ref{eq:GenTCPLinearSmallBuffer}),   and
obtain
\begin{align}
  \myi r_\mym e^{\myi\overline{\theta}_\mym(t)}\dot{\overline{\theta}}_\mym(t) =&\, - \Bigl( g(\weq{\mym})r_\mym e^{\myi\overline{\theta}_\mym(t)}\nonumber\\  &+\frac{b_\mym}{n_\mym} r_\mym e^{\myi\overline{\theta}_\mym(t-\tau_\mym)} \Bigr)\frac{i(\weq{\mym})}{\tau_\mym},\label{eq:GenTCPthetaSmall}
\end{align}
where  $\overline{\theta}_\mym(t)=\omega_\mym t+\phi_\mym$, $\omega_\mym$ is the intrinsic frequency of the $\mym^{th}$ set of flows and $\phi_\mym$ its phase. From
\eqref{eq:GenTCPthetaSmall} we get
\[
\begin{aligned}
g(\weq{\mym}) + \frac{b_\mym}{n_\mym}\cos(\omega_\mym\tau_\mym) &=0, \notag \\ 
\omega_\mym-\frac{i(\weq{\mym})}{\tau_\mym}b_\mym\sin(\omega_\mym\tau_\mym) &=0 \notag.
\end{aligned}
\]
On simplification, we obtain
\begin{equation}
\omega_\mym =\frac{i(\weq{\mym})}{\tau_\mym}\sqrt{\left(\frac{b_\mym}{n_\mym}\right)^2 - g^2\big(\weq{\mym}\big)}\,\,. \label{eq:GenTCPFreqSmall} 
\end{equation}
Substituting the functional forms for $i(\w{})$ and $d(\w{})$ for Compound, Reno and Illinois (see Table~\ref{table:windowfunctions}), we compute the 
intrinsic frequencies. 
The expressions for the intrinsic frequencies, in the case of small Drop-Tail buffers and smooth traffic $(n_\mym=1)$,  are listed in Table~\ref{table:small_intrinsic_frequencies}.\\

\noindent\textit{Intermediate buffers and Drop-Tail}\\ 
In an intermediate buffer regime, the buffers are sized as $B=c*\overline{RTT}/\sqrt{N}$, where $c$ 
is the link capacity, $\overline{RTT}$ is the average round-trip time of the flows, and $N$ is the number of flows~\cite{mckoewn}. 
In practice, $\overline{RTT}$ is fixed at $250$ ms. We now outline the model for the packet loss probability with smooth and bursty traffic.\\
 
\noindent\textit{Smooth traffic:} In this scenario, when the average arrival rate to the queue $x_{\mym}(t)=\w{\mym}(t)/\tau<c'_{\mym},$ the packet drop probability is zero, when the number of TCP flows is large. On the other hand, when $x_{\mym}(t)>c'_{\mym},$ a fluid model for intermediate buffers with Drop-Tail is \cite{rw}  
\begin{equation}
 p_\mym\big(\w{\mym}(t)\big) = \left(\frac{\w{\mym}(t)-c'_\mym\tau_\mym}{\w{\mym}(t)}\right)^+,
\label{eqn:edgeIntermediate}
\end{equation}
where $(z)^+$ is defined as $\max(z,0)$, $c'_\mym$ is the link capacity per flow, $\tau_\mym$ is the round-trip time and $x_\mym(t)$ is the rate, where $\w{\mym}(t)=x_\mym(t)\tau_\mym$.

In a deterministic fluid model, \eqref{eqn:edgeIntermediate} has the interpretation
of the fraction of fluid lost when the arrival rate exceeds capacity~\cite{srikant}.\\
 
\noindent\textit{Bursty traffic:} In this scenario, a batch of size $w(t)$ experiences a drop at time instant $t$ if the empty space in the buffer is less than $w(t).$ This leads to the packet drop probability at edge routers as \cite{rw,Wischik}
 \begin{equation}
 p_\mym\big(\w{\mym}(t)\big) = \frac{1}{\w{\mym}(t)}\left(\frac{\w{\mym}^{\w{\mym}(t)}(t)-\left(c'_\mym\right)^{\w{\mym}(t)}\tau_\mym^{\w{\mym}(t)}}{\w{\mym}^{\w{\mym}(t)}(t)}\right)^+,
\label{eqn:edgeIntermediate_bursty}
\end{equation}
For the sake of analysis, we again assume that the batch size is constant, and is denoted by $n_{\mym};$ $m=1,2.$ This yields the packet drop probability at the edge routers as  
\begin{equation}
 p_\mym\big(\w{\mym}(t)\big) = \frac{1}{n_{\mym}}\left(\frac{\w{\mym}^{n_{\mym}}(t)-\left(c'_\mym\right)^{n_{\mym}}\tau_\mym^{n_{\mym}}}{\w{\mym}^{n_{\mym}}(t)}\right)^+,
\label{eqn:edgeIntermediate_burstyconstant}
\end{equation}

\begin{table}[t]
\centering
\newcolumntype{T}{>{\flushleft\arraybackslash} m{2.5cm} }
\newcolumntype{S}{>{\flushright\arraybackslash}  m{3.5cm} }
\newcolumntype{F}{ m{0.95cm} }
 \begin{tabular}{TSF}
  \hline\\[-3ex]
  TCP  &   \centering \textit{Intrinsic frequency ($\omega_\mym$):\\ small buffers  }  &\\[7ex]
  \hline
  \hline
  
   \textit{Compound}  & 
\begin{tiny}$\displaystyle  \frac{\alpha{\weq{\mym}}^{k-1}}{\tau_\mym}\sqrt{b_\mym^2-(k-2)^2\big(1-p_\mym(\weq{\mym})\big)^2} $\end{tiny}  \\[2ex]
 
   \textit{Reno}  
   &$\displaystyle \frac{1}{\weq{\mym}\tau_\mym}\sqrt{b_\mym^2-4\big(1-p_\mym(\weq{\mym})\big)^2} $ \\[2ex]
    \textit{Illinois} 
    &$\displaystyle \frac{\alpha_{max}}{\weq{\mym}\tau_\mym}\sqrt{b_\mym^2- 4\big(1-p_\mym(\weq{\mym})\big)^2}$  \\[3.5ex]

  \hline
  \hline
 \end{tabular}
 \vspace{8pt}
\caption{ Intrinsic frequencies of the transport protocols with small Drop-Tail buffers and smooth traffic.}
\label{table:small_intrinsic_frequencies}
\end{table} 

We now analyse \eqref{eq:tcpeq} coupled with \eqref{eqn:edgeIntermediate_burstyconstant}. Let   $\w{\mym}(t)=\weq{\mym}+\varDelta{}\w{\mym}(t)$ and linearising system~\eqref{eq:tcpeq} and~\eqref{eqn:edgeIntermediate_burstyconstant}, about the equilibrium~$\w{\mym}^*$, gives

\begin{small}
\begin{align}
 \varDelta{}\dot{\w{}}_\mym(t) =&\,\,- \frac{g(\weq{\mym})i(\weq{\mym})}{\tau_\mym}\varDelta{}\w{\mym}(t)\nonumber\\&-\frac{\bigl( i(\weq{\mym})  + d(\weq{\mym})\bigr)\left(c'_\mym\right)^{n_\mym}\tau_{\mym}^{n_{\mym}-1}}{\left(\weq{\mym}\right)^{n_{\mym}}} \varDelta{}\w{\mym}(t-\tau_\mym). \label{eq:GenTCPLinInter}
\end{align}
\end{small}
%

%
%

\noindent\textit{Intrinsic frequencies:} 
As considered earlier, we assume that loss of local stability would give rise to limit cycles. 
Again, we approximate the limit cycles with harmonic oscillations.
To obtain their frequency, we substitute $\varDelta{}\w{\mym}(t) = r_\mym e^{\myi\overline{\theta}_\mym(t)}$ in \eqref{eq:GenTCPLinInter}   to obtain

\begin{small}
\begin{align}
   \myi r_\mym e^{\myi\overline{\theta}_\mym(t)}\dot{\overline{\theta}}_\mym(t) =\,\, - \frac{g(\weq{\mym})i(\weq{\mym})}{\tau_\mym}r_\mym e^{\myi\overline{\theta}_\mym(t)}\nonumber\\
   -\frac{\bigl( i(\weq{\mym})  + d(\weq{\mym})\bigr)\left(c'_\mym\right)^{n_\mym}\tau_{\mym}^{n_{\mym}-1}}{\left(\weq{\mym}\right)^{n_{\mym}}}r_\mym e^{\myi\overline{\theta}_\mym(t-\tau_\mym)} ,\label{eq:thetaGeneralInter}
 \end{align}
\end{small}

\noindent{}where  $\overline{\theta}_\mym(t)=\omega_\mym t+\phi_\mym$, $\omega_\mym$ is the intrinsic frequency of the $\mym^{th}$ set of flows and $\phi_\mym$ its phase. From
(\ref{eq:thetaGeneralInter}),
\[
\begin{scriptsize}
\begin{aligned}
\frac{g(\weq{\mym})i(\weq{\mym})}{\tau_{\mym}} + \frac{\bigl( i(\weq{\mym})  + d(\weq{\mym})\bigr)\left(c'_\mym\right)^{n_\mym}\tau_{\mym}^{n_{\mym}-1}}{\left(\weq{\mym}\right)^{n_{\mym}}}\cos(\omega_{\mym}\tau_{\mym}) &=0, \notag \\ 
\omega_{\mym}- \frac{\bigl( i(\weq{\mym})  + d(\weq{\mym})\bigr)\left(c'_\mym\right)^{n_\mym}\tau_{\mym}^{n_{\mym}-1}}{\left(\weq{\mym}\right)^{n_{\mym}}}\sin(\omega_{\mym}\tau_{\mym}) &=0 \notag,
\end{aligned}
\end{scriptsize}
\]
which gives
\begin{scriptsize}
\begin{equation}
  \omega_{\mym} =\sqrt{  \left(\frac{\bigl( i(\weq{\mym})  + d(\weq{\mym})\bigr)\left(c'_\mym\right)^{n_\mym}\tau_{\mym}^{n_{\mym}-1}}{\left(\weq{\mym}\right)^{n_{\mym}}}\right)^2 - 
   \left(\frac{i(\weq{\mym})g(\weq{\mym})}{\tau_{\mym}}\right)^2}. \label{eq:GenTCPFreqInter}
\end{equation}
   \end{scriptsize}

Using the functional forms for $i(\w{})$ and $d(\w{})$ listed in Table~\ref{table:windowfunctions}, we compute the 
intrinsic frequencies for the three protocols. 
The intrinsic frequencies in case of intermediate Drop-Tail buffers with smooth traffic $(n_{\mym}=1)$ are outlined in Table~\ref{table:inter_intrinsic_frequencies}.

%

\begin{table}
\centering
\newcolumntype{A}{>{\arraybackslash} m{2.5cm} }
\newcolumntype{B}{>{\flushright\arraybackslash} m{3.2cm} }
\newcolumntype{C}{>{\flushright} m{3.5cm} }
\newcolumntype{D}{>{\arraybackslash} m{0.05ex} }
 \begin{tabular}{ACD}
  \hline
  \\[-3ex]
   TCP
  & \centering \textit{Intrinsic frequency ($\omega_{\mym}$):\\ intermediate buffers  }   &\\[5ex]
  \hline
  \hline
   \textit{Compound}  & 
 $\displaystyle \frac{\beta \weq{\mym}}{\tau_{\mym}}\sqrt{1-(k-2)^2p_{\mym}^2(\weq{\mym})} $  & \\[6ex]
 
   \textit{Reno}  &$\displaystyle \frac{ \weq{\mym}}{2\tau_{\mym}}\sqrt{1-4p_{\mym}^2(\weq{\mym})}$  &  \\[6ex]
    \textit{Illinois}    &$\displaystyle \frac{\beta_{min} \weq{\mym}}{\tau_{\mym}}\sqrt{1-4p_{\mym}^2(\weq{\mym})}$  &\\[9ex]

  \hline
  \hline
 \end{tabular}
 \vspace{8pt}
\caption{ Intrinsic frequencies of the transport protocols with intermediate Drop-Tail buffers and smooth traffic.}
\label{table:inter_intrinsic_frequencies}
\end{table} 

Now that we have described the underlying fluid models, we go on to analyse their coupled dynamics, in Section \ref{section:coupled}, and the phenomena of synchronisation, 
in Section \ref{section:synchronisation}.


\newcommand{\corep}{p_{c}} 
\section{Coupled dynamics}
\label{section:coupled}
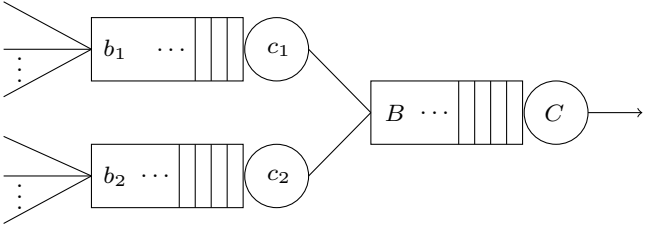
\begin{figure}[t]
\centering
\begin{tikzpicture}[scale=1.05, transform shape]
    \draw[-] (0.0,2.6) -- (1.1,2.0);
    \draw[-] (0.0,2.0) -- (1.1,2.0);
    \draw[-] (0.0,1.4) -- (1.1,2.0);
    \node at (0.2,1.85) {$\vdots$};
    \draw (1.1,1.6) rectangle(3.0,2.4);
    \draw (2.8,1.6) -- (2.8,2.4);
    \draw (2.6,1.6) -- (2.6,2.4);
    \draw (2.4,1.6) -- (2.4,2.4);   
    \draw (3.42,2.0) circle(0.4cm);
    \node at (3.45,2.0) {$c_1$};
    \node at (1.4,2.0) {$b_1$};
    \node at (2.1,2.0) {$\ldots$};
    \draw[-] (3.82,2.0) -- (4.6,1.2);
    
    \draw[-] (0.0,-0.2) -- (1.1,0.4);
    \node at (0.2,0.25) {$\vdots$};
    \draw[-] (0.0,0.9) -- (1.1,0.4);
    \draw[-] (0.0,0.4) -- (1.1,0.4);
    \draw (1.1,0) rectangle(3.0,0.8);
    \draw (2.8,0) -- (2.8,0.8);
    \draw (2.6,0) -- (2.6,0.8);
    \draw (2.4,0) -- (2.4,0.8);
    \draw (2.2,0) -- (2.2,0.8);
    \node at (1.4,0.4) {$b_2$};
    \node at (3.45,0.4) {$c_2$};
    \node at (1.9,0.4) {$\ldots$};
    \draw (3.42,0.4) circle(0.4cm);
    \draw[-] (3.82,0.4) -- (4.6,1.2);

    \draw (4.6,0.8) rectangle(6.5,1.6);
    \draw (6.3,0.8) -- (6.3,1.6);
    \draw (6.1,0.8) -- (6.1,1.6);
    \draw (5.9,0.8) -- (5.9,1.6);
    \draw (5.7,0.8) -- (5.7,1.6);
    \node at (4.9,1.2) {$B$};
    \node at (6.9,1.2) {$C$};
    \node at (5.4,1.2) {$\ldots$};
    \draw (6.92,1.2) circle(0.4cm);
    \draw[->] (7.32,1.2) -- (8.0,1.2); 
\end{tikzpicture}
\caption{ \small{ \textit{System Topology}: Two sets of  TCP 
flows from distinct edge routers, with different feedback delays, flowing through a common core router. The edge routers have buffer sizes $b_1$ and $b_2$ and the core router has a buffer size of $B$. The edge routers have link capacities $c_1$ and $c_2$ and the core router has a link capacity of $C.$} }
\label{fig:coupled}
\end{figure}

We consider the case where the two sets of TCP flows, from different edge routers, flow through a common core router.
The core router has link capacity $C$, which is less than the sum of the edge router capacities.
Thus the core router forms a bottleneck. Let the link capacity per flow for the core router be $C'$. The queue policy at the core is Drop-Tail. See Figure~\ref{fig:coupled} for the system topology.
 We provide some motivation for the topology that we study. This topology represents a network of three bottlenecked queues; i.e. two edge routers and a core router. Thus it goes beyond a simple single-bottleneck scenario, and is able to account for some network effects. The existence and stability of synchronised states, with TCP and a Drop-Tail queue management policy, is analysed in such a multi-bottlenecked topology.
The dynamical representation for this system for a generalised TCP is 
%
%
%
%

\begin{scriptsize}
\begin{align}
  &\dot{\w{}}_{\mym}(t) =\biggl( i\big(\w{\mym}(t)\big)\Bigl(1 - \big(p_{\mym}(t-\tau_{\mym}) +  \corep(t,\tau_1,\tau_2)\big) \Bigr) \label{eq:GenTCPCoupled}   \\
  &- d\bigl(\w{\mym}(t)\bigr)  \bigl(p_{\mym}(t-\tau_{\mym}) +  \corep(t,\tau_1,\tau_2) \bigr) \!\biggr)\!\dfrac{\w{\mym}(t-\tau_{\mym})}{\tau_{\mym}}; \mym=1,2, \notag
\end{align}
\end{scriptsize}
\hspace{-4pt}where $p_{\mym}(t)$ is the packet loss probability in the $\mym^{th}$ \textit{edge} router and $\corep(t,\tau_1,\tau_2)$ is the packet loss probability in the \textit{core} router.
\subsection{Small buffers and Drop-Tail}
From the fluid model approximation for small buffers and Drop-Tail (see Section \ref{subsection:FluidmodelGeneralisedTCP}), the packet loss model at the core router with bursty incoming traffic is
\begin{equation}\small
 \corep(t,\tau_1,\tau_2)=\frac{1}{n_c}\bigg(\frac{\sum_{\myj=1}^{2}\w{\myj}(t-\tau_\myj )/\tau_\myj }{\widetilde{C}}\bigg)^{\frac{B}{n_c}},
\label{eq:corerouterSmall}
\end{equation}
where $B$ is the buffer size, $n_c$ is the burstiness parameter and $\widetilde{C}=2C'.$ Here, $C'$ is the link capacity per flow of the core router. Note that when $n_c=1,$  we obtain the packet loss probability with smooth traffic.
%

%

%
We first substitute (\ref{eq:corerouterSmall}) and (\ref{eq:edgeSmall_burstyconstant}) into (\ref{eq:GenTCPCoupled}), and
assume that $\weqC{m}$ is the equilibrium of~\eqref{eq:GenTCPCoupled}. Thus, at equilibrium, we have the following packet loss probabilities

\begin{scriptsize}
\begin{equation*}\begin{aligned}
 p_{\mym}(\weqC{\mym}) &= \frac{1}{n_{\mym}}\bigg(\frac{\weqC{\mym}}{c'_{\mym}\tau_{\mym}}\bigg)^{\frac{b_{\mym}}{n_\mym}} \,\,\text{and} & \corep\big(\weqC{1},\weqC{2}\big)&= \frac{1}{n_c}\Biggl(\frac{\sum_{\myj=1}^{2}\weqC{\myj}/\tau_\myj }{\widetilde{C}}\Biggr)^{\frac{B}{n_c}}.
\end{aligned}\end{equation*}
\end{scriptsize}
Let $\w{m}(t)=\weqC{m}+\dw{m}(t)$, and linearising the non-linear system \eqref{eq:GenTCPCoupled}, about the equilibrium $\weqC{m}$, yields

\begin{scriptsize}
\begin{equation}\begin{aligned}
 \varDelta{}\dot{\w{}}_{\mym}(t)  =&\,\,  -\frac{i(\weqC{\mym}) }{\tau_{\mym}} \biggl( g(\weqC{\mym})\dw{\mym}(t)  + \frac{b_{\mym}}{n_{\mym}}\dw{\mym}(t-\tau_{\mym}) \!\biggr) \\ 
	&\,\,    - \Biggl(  \frac{B\weqC{\mym}/\tau_{\mym}}{n_c\sum_{\myj=1}^{2}\weqC{\myj}/\tau_\myj } 
	    \sum_{\myj=1}^{2}\frac{\dw{\myj}(t-\tau_\myj )}{\tau_\myj } \\
	&\,\, - \frac{b_{\mym}}{n_\mym}\frac{\dw{\mym}(t-\tau_{\mym})}{\tau_{\mym}}\Biggr)\frac{i\big(\weqC{\mym}\big)}{1+p_{\mym}\big(\weqC{\mym}\big)/\corep\big(\weqC{1},\weqC{2}\big)} \label{eq:linearSmallCoupledMod2}.
\end{aligned}\end{equation}
\end{scriptsize}

%
%
%
%
An underlying assumption in systems that are weakly coupled is that the relative positions of the oscillators would change slowly, compared to their motion around the limit cycle. Thus, the packet loss probability $\corep(.)$ is assumed to be small enough to ensure weak coupling in the system dynamics.    
We now aim to transform (\ref{eq:linearSmallCoupledMod2}) into a model for weakly coupled oscillators.
%
Hence $\weqC{\mym}\approx\weq{\mym}$, and

\begin{scriptsize}
\begin{equation}\begin{aligned}
 \varDelta{}\dot{\w{}}_{\mym}(t)  =&\,  -\frac{i(\weq{\mym}) }{\tau_{\mym}} \biggl( g(\weq{\mym})\dw{\mym}(t)  + \frac{b_{\mym}}{n_\mym}\dw{\mym}(t-\tau_{\mym}) \biggr) \\ 
	&\,   - \Biggl(  \frac{B\weq{\mym}/\tau_{\mym}}{n_c\sum_{\myj=1}^{2}\weq{\myj}/\tau_\myj } 
	    \sum_{\myj=1}^{2}\frac{\dw{\myj}(t-\tau_\myj )}{\tau_\myj } \\
	&\, - \frac{b_{\mym}}{n_m}\frac{\dw{\mym}(t-\tau_{\mym})}{\tau_{\mym}}\Biggr)\frac{i(\weq{\mym})}{1+p_{\mym}(\weq{\mym})/\corep(\weq{1},\weq{2})} \label{eq:linearSmallCoupledMod3}.
\end{aligned}\end{equation}
\end{scriptsize}

We substitute  $\varDelta{}\w{\mym}(t)=r_{\mym} e^{\myi\theta_{\mym}(t)}$ into (\ref{eq:linearSmallCoupledMod3}) to obtain
\begin{equation*}\begin{aligned}
 &\myi r_{\mym}e^{\myi\theta_{\mym}(t)}\dot{\theta}_{\mym}(t)   =  -\frac{i(\weq{\mym}) }{\tau_{\mym}} \bigg( g(\weq{\mym})r_{\mym} e^{\myi\theta_{\mym}(t)}  \\ 
	&+ \frac{b_{\mym}}{n_m}r_{\mym} e^{\myi\theta_{\mym}(t-\tau_{\mym})} \bigg)   -  \Biggl(  \frac{B\weq{\mym}/\tau_{\mym}}{n_c\sum_{\myj=1}^{2}\weq{\myj}/\tau_\myj } 
	\sum_{\myj=1}^{2}\frac{r_\myj e^{\myi\theta_\myj(t-\tau_\myj )}}{\tau_\myj } \\
	&-  \frac{b_{\mym}}{n_m}\frac{r_{\mym} e^{\myi\theta_{\mym}(t-\tau_{\mym})}}{\tau_{\mym}}  \Biggr)\frac{i(\weq{\mym})}{1+p_{\mym}(\weq{\mym})/\corep(\weq{1},\weq{2})}.
\end{aligned}\end{equation*}
Simplifying the above equation, we get
\begin{small}
\begin{align}
 \myi\dot{\theta}_{\mym}(t)  =&\,\, - \frac{i(\weq{\mym}) }{\tau_{\mym}} \bigg( g(\weq{\mym}) + \frac{b_{\mym}}{n_\mym} e^{\myi\theta_{\mym}(t-\tau_{\mym})-\myi\theta_{\mym}(t)} \bigg)  \notag \\ 
	&\,\,   - \Biggl( \frac{B\weq{\mym}/\tau_{\mym}}{n_c\sum_{\myj=1}^{2}\weq{\myj}/\tau_\myj }\sum_{\myj=1}^{2}\frac{r_\myj e^{\myi\theta_\myj(t-\tau_\myj )-\myi\theta_{\mym}(t)}}{\tau_\myj r_{\mym}} \label{eq:linearSmallCoupledMod1} \\
	&\,\,-  \frac{b_{\mym}}{n_\mym}\frac{e^{\myi\theta_{\mym}(t-\tau_{\mym})-\myi\theta_{\mym}(t)}}{\tau_{\mym}} \Biggr)\frac{i(\weq{\mym})}{1+p_{\mym}(\weq{\mym})/\corep(\weq{1},\weq{2})} \notag .
\end{align}
\end{small}

\vspace{-1ex}
\noindent{}Since the coupling is weak, at the fast time scale, the
oscillators behave almost independently. So $\theta_{\mym}\approx\overline{\theta}_{\mym}$. Thus, (\ref{eq:linearSmallCoupledMod1}) can be written as

\begin{scriptsize}
\begin{equation*}\begin{aligned}
 \myi\dot{\theta}_{\mym}(t)  =&\,\,  -\frac{i(\weq{\mym}) }{\tau_{\mym}} \bigg( g(\weq{\mym})  + \frac{b_{\mym}}{n_\mym} e^{\myi\overline{\theta}_{\mym}(t-\tau_{\mym})-\myi\overline{\theta}_{\mym}(t)} \bigg)   \\
	&\,\, - \Biggl( \frac{B\weq{\mym}/\tau_{\mym}}{n_c\sum_{\myj=1}^{2}\weq{\myj}/\tau_\myj }\sum_{\myj=1}^{2}\frac{r_\myj e^{\myi\theta_\myj(t-\tau_\myj )-\myi\theta_{\mym}(t)}}{\tau_\myj r_{\mym}} \\
	&\,\, -  \frac{b_{\mym}}{n_\mym}\frac{e^{\myi\theta_{\mym}(t-\tau_{\mym})-\myi\theta_{\mym}(t)}}{\tau_{\mym}} \Biggr)\frac{i(\weq{\mym})}{1+p_{\mym}(\weq{\mym})/\corep(\weq{1},\weq{2})}.
\end{aligned}\end{equation*}
\end{scriptsize}
Substituting (\ref{eq:GenTCPthetaSmall}) into the above equation gives

\begin{scriptsize}
\begin{equation*}\begin{aligned}
 \myi\dot{\theta}_{\mym}(t)  =&\,\,  \myi\omega_{\mym}   - \Biggl( \frac{B\weq{\mym}/\tau_{\mym}}{n_c\sum_{\myj=1}^{2}\weq{\myj}/\tau_\myj }\sum_{\myj=1}^{2}\frac{r_\myj e^{\myi\theta_\myj(t-\tau_\myj )-\myi\theta_{\mym}(t)}}{\tau_\myj r_{\mym}} \\
 &-  \frac{b_{\mym}}{n_\mym}\frac{e^{\myi\theta_{\mym}(t-\tau_{\mym})-\myi\theta_{\mym}(t)}}{\tau_{\mym}} \Biggr)
 \frac{i(\weq{\mym})}{1+p_{\mym}(\weq{\mym})/\corep(\weq{1},\weq{2})}.
\end{aligned}\end{equation*}
\end{scriptsize}
Equating the imaginary part of the above equation, we obtain the following model for coupled oscillators

\vspace{-2ex}
{\small\begin{align}
 &\dot{\theta}_{\mym}(t)  =  \omega_{\mym}  - \Biggl( \frac{B\weq{\mym}/\tau_{\mym}}{n_c\sum_{\myj=1}^{2}\weq{\myj}/\tau_\myj }\sum_{\myj=1}^{2}\frac{r_\myj \sin\big(\theta_\myj(t-\tau_\myj )-\theta_{\mym}(t)\big)}{\tau_\myj r_{\mym}} \notag \\
 &\!\!\!-  \frac{b_{\mym}}{n_\mym}\frac{\sin\big(\theta_{\mym}(t-\tau_{\mym})-\theta_{\mym}(t)\big)}{\tau_{\mym}} \Biggr)\frac{i(\weq{\mym})}{1+p_{\mym}(\weq{\mym})/\corep(\weq{1},\weq{2})}   \label{eq:GenTCPlinearSmallCoupledModFinal}.
\end{align}}
%
%
%
%
\hspace{-3pt}For now we focus on the case where both sets of flows, with respect to the core router, are equally coupled.  \\

\noindent\textit{Equally coupled TCP flows} \\
Let  $b_1=b_2=b$, $c'_1=c'_2=c$, $n_1=n_2=n_e$ and $\tau_1 \approx \tau_2 = \tau$. Then, $\weq{\mym} \,\dot{=}\,\, \w{}^*$ and $r_1 \approx r_2$.
Hence, (\ref{eq:GenTCPlinearSmallCoupledModFinal}) can be expressed as the following model for coupled oscillators
\begin{equation}\begin{aligned}
\dot{\theta}_{\mym}(t) =&\,\, \omega_{\mym} - \frac{B}{2n_c}K_s\!\sum_{\myj=1}^{2}\sin\Bigl(\theta_\myj(t-\tau_\myj )\!-\theta_{\mym}(t)\Bigr)\\ &+ \frac{b}{n_e}K_s\sin\Bigl(\theta_{\mym}(t-\tau_{\mym})-\theta_{\mym}(t)\Bigr) , \label{eq:linearSmallEquallyCoupled}
\end{aligned}\end{equation}
where the \emph{small buffer coupling strength}, $K_s$ is
\begin{equation}\small\begin{aligned}
 K_s   = \frac{i(\weq{})}{\tau\big(1+p(\weq{})/\corep(\weq{},\weq{})\big)}. \label{eq:couplingStrengthSmall}                
\end{aligned}\end{equation}
%

Note that when the burstiness parameters are set to 1, we obtain the expressions corresponding to smooth traffic. We now consider the intermediate router buffer sizing regime for our analysis.
\subsection{Intermediate buffers and Drop-Tail}
From the fluid model approximation for intermediate buffers and Drop-Tail given in Section \ref{subsection:FluidmodelGeneralisedTCP}, the packet loss model at the core router is
\begin{equation}\small
 \corep(t,\tau_1,\tau_2) =\frac{1}{n_c}\biggl(\frac{\big(\sum_{\myj=1}^{2}\w{\myj}(t-\tau_\myj )/\tau_\myj  \big)^{n_c} - \left(\widetilde{C}\right)^{n_c}}{\left(\sum_{\myj=1}^{2}\w{\myj}(t-\tau_\myj )/\tau_\myj\right)^{n_c} } 
\biggr)^+, \label{eq:corerouterIntermediate}
\end{equation}  
where $\widetilde{C}=2C'$, and $C'$ is the link capacity  per flow of the core router.

We first substitute (\ref{eq:corerouterIntermediate}) and (\ref{eqn:edgeIntermediate}) into the coupled TCP model 
(\ref{eq:GenTCPCoupled}), and let $\w{m}(t)=\weqC{m}+\dw{m}(t)$, where $\weqC{m}$ is the equilibrium of (\ref{eq:GenTCPCoupled}). 
Then, linearising the non-linear system~\eqref{eq:GenTCPCoupled},  we get

\vspace{-1.5ex}
\begin{align}
  \varDelta{}\dot{\w{}}_{\mym}(t) =  &-\frac{i(\weqC{\mym})g(\weqC{\mym})}{\tau_{\mym}}\dw{\mym}(t) -\bigl( i(\weqC{\mym})  + d(\weqC{\mym})\bigr)  \notag\\ 
		    & \times\Biggl(  \frac{\left(c'_{\mym}\right)^{n_\mym}\tau_{m}^{n_m-1}}{\left(\weqC{\mym}\right)^{n_m}} \dw{\mym}(t-\tau_{\mym})\notag\\
		    &  +  \frac{\left(\widetilde{C}\right)^{n_c}\weqC{\mym}/\tau_{\mym}}{\big(\sum_{\myj=1}^{2}\weqC{\myj}/\tau_\myj \big)^{n_c+1}}\sum_{\myj=1}^{2}\frac{\dw{\myj}(t-\tau_\myj )}{\tau_\myj } \Biggr). \label{eq:GenTCPlinearIntermediateCoupledMod2}    
\end{align}
 
Our objective is to transform system (\ref{eq:GenTCPlinearIntermediateCoupledMod2}) into a model for coupled oscillators.
As earlier, we assume that $\corep(.)$ as defined in (\ref{eq:corerouterIntermediate}) is small, thus the \emph{coupling is weak}.
Hence $\weqC{\mym}\approx\weq{\mym}$ and (\ref{eq:GenTCPlinearIntermediateCoupledMod2}) becomes
%
%
%

\vspace{-1.5ex}\begin{align}
  \varDelta{}\dot{\w{}}_{\mym}(t) &=  -\frac{i(\weq{\mym})g(\weq{\mym})}{\tau_{\mym}}\dw{\mym}(t) -\bigl( i(\weq{\mym})  + d(\weq{\mym})\bigr)  \notag \\ 
		    & \times\Biggl( \frac{ \left(c'_{\mym}\right)^{n_\mym}\tau_{\mym}^{n_m-1}}{\left(\weq{\mym}\right)^{n_\mym}} \dw{\mym}(t-\tau_{\mym})\notag\\
		    & +  \frac{\left(\widetilde{C}\right)^{n_c}\weq{\mym}/\tau_{\mym}}{\big(\sum_{\myj=1}^{2}\weqC{\myj}/\tau_\myj \big)^{n_c+1}}\sum_{\myj=1}^{2}\frac{\dw{\myj}(t-\tau_\myj )}{\tau_\myj } \Biggr).
		     \label{eq:GenTCPlinearIntermediateCoupledMod3}  
\end{align}
\hspace{-3.5pt}Substituting  $\varDelta{}\w{\mym}(t)=r_{\mym} e^{\myi\theta_{\mym}(t)}$ into (\ref{eq:GenTCPlinearIntermediateCoupledMod3}), we obtain
\vspace{0ex}
\begin{scriptsize}\begin{align*}
  \myi r_{\mym}e^{\myi\theta_{\mym}(t)}\dot{\theta}_{\mym}(t) = & -\frac{i(\weq{\mym})g(\weq{\mym})}{\tau_{\mym}}r_{\mym} e^{\myi\theta_{\mym}(t)}  -\bigl( i(\weq{\mym})  + d(\weq{\mym})\bigr)   \notag \\
  & \times\Biggl( \frac{\left(c'_{\mym}\right)^{n_\mym}\tau_{\mym}^{n_m-1}}{\left(\weq{\mym}\right)^{n_\mym}} r_{\mym} e^{\myi\theta_{\mym}(t-\tau_{\mym})}\notag\\
		   & +  \frac{\left(\widetilde{C}\right)^{n_c}\weq{\mym}/\tau_{\mym}}{\big(\sum_{\myj=1}^{2}\weq{\myj}/\tau_\myj \big)^{n_c+1}}\sum_{\myj=1}^{2}\frac{r_\myj  e^{\myi\theta_\myj(t-\tau_\myj )}}{\tau_\myj }\Biggr).  
\end{align*}
\end{scriptsize}
\hspace{-3pt}Simplifying the above equation, we get
%
\vspace{0ex}
\begin{scriptsize}\begin{align}
  \myi\dot{\theta}_{\mym}(t) &= -\frac{i(\weq{\mym})g(\weq{\mym})}{\tau_{\mym}} \notag\\
  &- \frac{\bigl( i(\weq{\mym})  + d(\weq{\mym})\bigr)\left(c'_{\mym}\right)^{n_\mym}\tau_{\mym}^{n_m-1}}{\left(\weq{\mym}\right)^{n_\mym}} e^{\myi\theta_{\mym}(t-\tau_{\mym})-\myi\theta_{\mym}(t)}  \notag \\
  &-  \frac{\big( i(\weq{\mym})+d(\weq{\mym}) \big)\left(\widetilde{C}\right)^{n_c}\weq{\mym}/\tau_{\mym}}{\big(\sum_{\myj=1}^{2}\weq{\myj}/\tau_\myj \big)^{n_c+1}}\sum_{\myj=1}^{2}\frac{r_\myj e^{\myi\theta_\myj(t-\tau_\myj )-\myi\theta_{\mym}(t)}}{\tau_\myj r_{\mym}}. \label{eq:linearIntermediateCoupledMod4}  
\end{align}
\end{scriptsize}

\vspace{-1ex}\noindent{}Since the coupling is weak, at the fast time scale, the
oscillators behave almost independently. So  $\theta_{\mym}\approx\overline{\theta}_{\mym}$. Thus, (\ref{eq:linearIntermediateCoupledMod4}) can now be written as
%

\begin{scriptsize}
\begin{align}
  \myi\dot{\theta}_{\mym}(t) = &-\frac{i(\weq{\mym})g(\weq{\mym})}{\tau_{\mym}}\notag\\
  & - \frac{\bigl( i(\weq{\mym})  + d(\weq{\mym})\bigr)\left(c'_{\mym}\right)^{n_\mym}\tau_{\mym}^{n_m-1}}{\left(\weq{\mym}\right)^{n_\mym}} e^{\myi\overline{\theta}_{\mym}(t-\tau_{\mym})-\myi\overline{\theta}_{\mym}(t)}  \notag \\
  &-  \frac{\big( i(\weq{\mym})+d(\weq{\mym}) \big)\left(\widetilde{C}\right)^{n_c}\weq{\mym}/\tau_{\mym}}{\big(\sum_{\myj=1}^{2}\weq{\myj}/\tau_\myj \big)^{n_c+1}}\sum_{\myj=1}^{2}\frac{r_\myj e^{\myi\theta_\myj(t-\tau_\myj )-\myi\theta_{\mym}(t)}}{\tau_\myj r_{\mym}}.
\end{align}
\end{scriptsize}
Substituting (\ref{eq:thetaGeneralInter}) into the above equation gives
%
\begin{scriptsize}
\begin{equation*}\begin{aligned}
  \myi\dot{\theta}_{\mym}(t) &= \myi\omega_{\mym}-  \frac{\big( i(\weq{\mym})+d(\weq{\mym}) \big)\left(\widetilde{C}\right)^{n_c}\weq{\mym}/\tau_{\mym}}{\big(\sum_{\myj=1}^{2}\weq{\myj}/\tau_\myj \big)^{n_c+1}}\sum_{\myj=1}^{2}\frac{r_\myj  e^{\myi\theta_\myj(t-\tau_\myj )-\myi\theta_{\mym}(t)}}{\tau_\myj r_{\mym}}.   
\end{aligned}\end{equation*}
\end{scriptsize}
By equating the imaginary part of the above equation, we obtain the following model for
coupled oscillators
%
\begin{scriptsize}
\begin{equation}\begin{aligned}
  &\hspace{-2mm}\dot{\theta}_{\mym}(t) = \omega_{\mym}- \frac{\big( i(\weq{\mym})+d(\weq{\mym}) \big)\left(\widetilde{C}\right)^{n_c}\weq{\mym}/\tau_{\mym}}{\big(\sum_{\myj=1}^{2}\weq{\myj}/\tau_\myj \big)^{n_c+1}}
  \sum_{\myj=1}^{2}\frac{r_\myj  \sin\bigl( \theta_\myj(t-\tau_\myj )- \theta_{\mym}(t)\bigr)}{\tau_\myj r_{\mym}}. \label{eq:GenTCPlinearInterCoupledModFinal}
\end{aligned}\end{equation} 
\end{scriptsize}

\noindent\textit{Equally coupled TCP flows}\\
Let  $c'_1 = c'_2 = c$, $n_1=n_2=n_e$ 
and $\tau_1 \approx \tau_2 = \tau$. Then $\weq{\mym}\,\dot{=}\,\,\weq{}$ and $r_1 \approx r_2$.
Throughout the paper, $\doteq$ is used to denote `defined as'.
Thus (\ref{eq:GenTCPlinearInterCoupledModFinal}) can be expressed as the following model for coupled oscillators
\begin{equation} \begin{aligned}
          \dot{\theta}_{\mym}(t) &= \omega_{\mym}  -
   K_i{\sum\limits_{j=1}^{2}}\sin\Bigl(\theta_j(t-\tau_j )-\theta_{\mym}(t)\Bigr), \label{eq:linearIntermediateEquallyCoupled}
\end{aligned}\end{equation}
where the \textit{intermediate buffer coupling strength} is
\begin{equation}
 K_i=\frac{\big(i(\weq{}) + d(\weq{})\big)\left(\widetilde{C}\right)^{n_c}}{2^{n_c+1}\left(\weq{}\right)^{n_c}\tau^{1-n_c}}.  \label{eq:couplingStrengthInter}
\end{equation}

Note that when the burstiness parameters are set to 1, we obtain results corresponding to smooth traffic. In the next section, we present the main analytical results of the paper. The results pertain to the phenomenon of \emph{synchronisation} in communication networks when TCP flows are coupled with Drop-Tail queues.
\section{Synchronisation}
\label{section:synchronisation}
We now derive the conditions for synchronisation of the small buffer coupled oscillators (\ref{eq:linearSmallEquallyCoupled}) and the intermediate buffer
coupled oscillators (\ref{eq:linearIntermediateEquallyCoupled}) respectively.

\subsection{Small buffers}
We outline the conditions under which the small buffer coupled oscillators model 
(\ref{eq:linearSmallEquallyCoupled}) would synchronise.\\

{\begin{table}[t]
\centering
\newcolumntype{A}{>{\flushleft\arraybackslash} m{2.4cm} }
\newcolumntype{B}{>{\flushright\arraybackslash} m{2.5cm} }
\newcolumntype{C}{>{\flushright\arraybackslash} m{2.9cm} }
\newcolumntype{D}{>{\arraybackslash} l }
\resizebox{8cm}{!}{ 
 \begin{tabular}{ABCD}
 \hline \\[-1ex]
     TCP  & \centering \textit{Coupling strength:\\ small buffer  ($K_s$)} 
  & \centering \textit{Coupling strength: \\ intermediate  buffer   ($K_i$)  }  & \\[4ex]
  \hline
  \hline \\[-3ex]
   \textit{Compound}  &  $\displaystyle  \frac{\alpha {\weq{}}^{k-1}}{\tau\left(1  + \frac{(\weq{}/c)^b}{(2\weq{}/\widetilde{C})^B} \right)}$  & 
 $\displaystyle \big( \alpha{\weq{}}^{k-2} + \beta \big)\frac{\widetilde{C}}{4} $   & \\[5ex]
   \textit{Reno} &  $\displaystyle  \frac{1}{\weq{}\tau\left(1  + \frac{(\weq{}/c)^b}{(2\weq{}/\widetilde{C})^B} \right)}$  
  & $\displaystyle \left( \frac{2}{{\weq{}}^2} + 1 \right)\frac{\widetilde{C}}{8}$ &  \\[5ex]
   \textit{Illinois} &  $\displaystyle \frac{\alpha_{max}}{\weq{}\tau\left(1  + \frac{(\weq{}/c)^b}{(2\weq{}/\widetilde{C})^B} \right)}$ 
   & $\displaystyle \left( \frac{1}{{\weq{}}^2} + \beta_{min} \right)\frac{\widetilde{C}}{4}$ &\\[5ex]

  \hline
  \hline
  
 \end{tabular}
 }
 \vspace{8pt}
\caption{Coupling strengths of the transport protocols in small and intermediate Drop-Tail buffers, with smooth traffic.}
\label{table:couplingStrengths}
\end{table}

\newtheorem{thm1}{Theorem}[subsection]
\begin{thm1}
Assume that the difference between round-trip delays is small 
for the small buffer coupled oscillators in (\ref{eq:linearSmallEquallyCoupled}), i.e., $\tau_1 \approx \tau_2 = \tau$.
Then $\omega_1 \approx \omega_2 = \omega$. If there exists  $\Omega>0$ and $\phi_0>0$   satisfying
\begin{equation}
K_s\frac{B}{n_c}\sin{\phi_0}\cos(\Omega{}\tau)=\omega_2-\omega_1, \label{eq:relOmega1}
\end{equation}
and
 \begin{equation}
 \Omega{}=\omega+K_s\left(\frac{B}{2n_c}-\frac{b}{n_e}\right)\sin(\Omega\tau)+\frac{K_sB}{2n_c}\sin\bigl(\Omega\tau+\phi_0\bigr), \label{eq:relOmega2}
\end{equation} 
then synchronisation between coupled oscillators (\ref{eq:linearSmallEquallyCoupled}) would exist with frequency  $\Omega$ and phase difference $\phi_0$.
The necessary and sufficient condition for local stability of this synchronised state is given by
\begin{equation}
 K_s\left(\frac{B}{n_c}-\frac{b}{n_e}\right)\cos(\Omega\tau) < 0.  \label{eq:couplingstability}
\end{equation}
\end{thm1}

\begin{proof}
Using the two relations in (\ref{eq:linearSmallEquallyCoupled}), we form the following phase difference equation

\vspace{-2ex}
\begin{scriptsize}
\begin{align}
 &\dot{\theta}_1(t)- \dot{\theta}_2(t) = \omega_1 - \omega_2 \notag\\& -K_s\biggl( \frac{B}{2n_c}-\frac{b}{n_e} \biggr)\biggl[ \sin\Bigl( \theta_1(t-\tau_1) - \theta_1(t) \Bigr) - \sin\Bigl( \theta_2(t-\tau_2) - \theta_2(t) \Bigr) \biggr]\notag\\ &\quad\,\,-\frac{K_sB}{2n_c} \biggl[ \sin\Bigl( \theta_2(t-\tau_2) - \theta_1(t) \Bigr) - \sin\Bigl( \theta_1(t-\tau_1) - \theta_2(t) \Bigr) \biggr]. \label{eq:linearSmallEquallyCoupledMod1} 
\end{align}
\end{scriptsize}

Define $\phi(t)=\theta_1(t)-\theta_2(t)$ as the phase difference between the
oscillators. Let $\phi_0$ be its steady state value obtained by letting $\dot{\phi}=0$. 
Then, since $\tau_1 \approx \tau_2$,
 \[ \sin\Bigl( \theta_1(t-\tau_1) - \theta_1(t) \Bigr) - \sin\Bigl(\theta_2(t-\tau_2) - \theta_2(t) \Bigr) = 0, \] 
and
 \[ \theta_1\big(t-\tau_1\big) - \theta_2(t) \approx \theta_2\big(t-\tau_2\big) - \theta_1(t) + 2\phi_0. \] 
From the steady state condition in (\ref{eq:linearSmallEquallyCoupledMod1}), we obtain
 \begin{displaymath}
 K_s\frac{B}{n_c}\sin{\phi_0}\cos\Bigl(\theta_2(t-\tau_2) - \theta_1(t) + \phi_0\Bigr) \approx \omega_2 - \omega_1,  
\end{displaymath}
which implies that $\theta_2\big(t - \tau_2\big) - \theta_1(t)$, $\theta_2\big(t - \tau_2 \big) - \theta_2(t)$,
 $\theta_1\big(t - \tau_1\big) - \theta_1(t)$, and $\theta_1\big(t-\tau_1\big) - \theta_2(t)$ are constant. Thus
 \begin{align}
\dot{\theta}_1(t) 	=& \,\,\, \dot{\theta}_2(t) \notag \\
		=&\,\, \omega_1 - K_s\left(\frac{B}{2n_c} - \frac{b}{n_e}\right)\sin\Bigl(\theta_1\big(t-\tau_1\big) - \theta_1(t)\Bigr) \\
		  &\quad\,\,\,   -\frac{K_sB}{2n_c} \sin\Bigl(\theta_2\big(t-\tau_2\big) - \theta_1(t) \Bigr) \notag \\
		=&\,\, \omega_2 -   \frac{K_sB}{2n_c} \sin\Bigl( \theta_1(t-\tau_1) - \theta_2(t) \Bigr) \\
		   &\quad\,\,\, -K_s\left(\frac{B}{2n_c} - \frac{b}{n_e}\right)\sin\Bigl(\theta_2\big(t-\tau_2\big) - \theta_2(t) \Bigr) \notag \\
		\dot{=}&\,\,  \Omega .\notag
\end{align} 
Hence from the steady state conditions,
\[\begin{aligned}
 \theta_1(t)  &=\Omega{}t, \qquad \theta_2(t) =\Omega{}t - \phi_0, \notag \\
 \Omega &= \omega + K_s\left(\frac{B}{2n_c}-\frac{b}{n_e}\right)\sin\bigl(\Omega{}\tau\bigr)+\frac{K_sB}{2}\sin\bigl(\Omega{}\tau+\phi_0\bigr),  \notag
\end{aligned}\]
and
\[ K_s
frac{B}{n_c}\sin{\phi_0}\cos(\Omega\tau) = \omega_2 - \omega_1.  \]
As $K_s$ increases, $\phi_0$ approaches zero and (\ref{eq:relOmega2}) becomes
\[ \Omega{}=\omega+K_s\left(\frac{B}{n_c}-\frac{b}{n_e}\right)\sin(\Omega\tau). \] 
The expression (\ref{eq:couplingstability}) then follows from the analysis in \cite{strogatzStability}.
\end{proof}

Note that the theorem provides necessary and sufficient conditions for the local stability of the synchronised state, and the condition for the existence of the state is simply sufficient.

\vspace{4pt}
\noindent\textit{Discussion:}

The following insights are worth recapitulating from~\cite{strogatz}: (i) the phenomena of synchronisation occurs spontaneously as the coupling strength crosses a certain critical value, and (ii) that synchronisation occurs even in the presence of distinct delays.           

As far as our system is concerned, it can be shown that synchronisation becomes more prominent as the core buffer size increases. Thus, as the thresholds of the Drop-Tail buffer sizes increase, the more likely we are to witness the onset of synchronised dynamics in the queue sizes. This insight remains consistent for smooth as well as bursty traffic. A similar qualitative insight has been obtained in \cite{debayani1,debayani2}, through a different style of analysis. 

We now discuss the stability of the synchronised state for the generalised TCP model.
Let $K_s \in \big(K_{c_s}, K_{u_s}\big)$, where $K_{c_s}$ is the critical coupling strength~\cite{strogatz},  be the range for which equations
(\ref{eq:relOmega1}) -- (\ref{eq:couplingstability}) are feasible. 
 Then,
\begin{enumerate}
 \item[a)] For $K_s < K_{c_s}$, there exists no $\Omega$ and $\phi_0$ for which~\eqref{eq:relOmega1} and~\eqref{eq:relOmega2} are satisfied~\cite{strogatz}.
 Thus synchronisation does not occur.
 \item[b)] For $K_{c_s} < K_s < K_{u_s}$, at least one synchronised state would exist and is locally stable. 
 So the phase difference $\phi_0$ between the oscillators is locked.
 \item[c)] As the coupling strength further increases, the stability condition~\eqref{eq:couplingstability} gets violated. Hence the synchronised state is unstable
 for $K_s > K_{u_s}$. 
\end{enumerate}

Note that in the synchronised range, $\phi_0$ decreases with an increase in the \emph{coupling strength} $K_s$. For large values of $K_s$, $\phi_0$ tends to zero, and the arguments presented in~\cite{hhtc} show that in this limit we get $\Omega \approx \pi/\tau$. Also observe that for large values of $K_s$,  the synchronising frequency is independent of the intrinsic frequencies.

Now consider the complex order parameter, $r(t)e^{\myi\psi(t)}\\= \frac{1}{2}\left(e^{\myi\theta_1(t)}+e^{\myi\theta_2(t)}\right)$. 
The radius $r(t)$ measures the phase coherence between the oscillators, and  the angle $\psi(t)$ is the average phase. 
When the coupled oscillators synchronise, we have $r(t)=\cos(\phi_0/2)$. 
Thus, as the coupling strength $K_s$ increases, $\phi_0$ decreases, which results in increased phase coherence. 
%
%
%

\subsection{Intermediate buffers}
The following result provides the conditions under which the coupled oscillators 
(\ref{eq:linearIntermediateEquallyCoupled}) of the intermediate buffer regime would synchronise.\\

\newtheorem{thm2}{Theorem}[subsection]
\begin{thm2}
Assume that the difference between round-trip delays is small 
for the intermediate buffer coupled oscillators in (\ref{eq:linearIntermediateEquallyCoupled}), i.e., $\tau_1 \approx \tau_2 = \tau$.
Then $\omega_1 \approx \omega_2 = \omega$. If there exists  $\Omega>0$ and $\phi_0>0$   satisfying
\begin{equation}
2K_i\sin{\phi_0}\cos(\Omega{}\tau)=\omega_2-\omega_1, \label{eq:relIntermediateOmega1}
\end{equation}
and
\begin{equation}
 \Omega{}=\omega+K_i\sin(\Omega\tau)+K_i\sin\bigl(\Omega\tau+\phi_0\bigr), \label{eq:relIntermediateOmega2}
\end{equation}
then synchronisation between the coupled oscillators (\ref{eq:linearIntermediateEquallyCoupled}) would exist with frequency  $\Omega$ and phase difference $\phi_0$.
The necessary and sufficient condition for local stability of this synchronised state is given by
\begin{equation}
 2K_i\cos(\Omega\tau) < 0.  \label{eq:couplingIntermediatestability}
\end{equation}
\end{thm2}
\begin{proof}
Using the two relations in (\ref{eq:linearIntermediateEquallyCoupled}), we form the phase difference equation

\vspace{-2ex}
{\small \begin{align}
 &\hspace{-1mm}\dot{\theta}_1(t)- \dot{\theta}_2(t) = \omega_1 -\omega_2 \notag \\
      &\hspace{0.5mm} -K_i\biggl[\sin\Bigl(\theta_1(t-\tau_1) - \theta_1(t) \Bigr) 
       - \sin\Bigl(\theta_2(t-\tau_2) - \theta_2(t) \Bigr)\biggr] \label{eq:linearIntermediateEquallyCoupledMod1} \\ 
       &\hspace{0.5mm} -K_i\biggl[ \sin\Bigl( \theta_2(t-\tau_2)- \theta_1(t) \Bigr)
	 - \sin\Bigl(\theta_1(t-\tau_1) - \theta_2(t) \Bigr) \biggr]. \notag	 
\end{align}}
\hspace{-4pt}Define $\phi(t)=\theta_1(t)-\theta_2(t)$ as the phase difference between the
oscillators. Let $\phi_0$ denote its steady state value obtained by setting
 $\dot{\phi}=0$. Then, since $\tau_1 \approx \tau_2$,
 \[ \sin\Bigl( \theta_1(t-\tau_1) - \theta_1(t) \Bigr) - \sin\Bigl(\theta_2(t-\tau_2) - \theta_2(t) \Bigr) = 0, \] 
and
 \[ \theta_1(t-\tau_1) - \theta_2(t) \approx \theta_2(t-\tau_2) - \theta_1(t) + 2\phi_0. \]
From the steady state condition in (\ref{eq:linearIntermediateEquallyCoupledMod1}), we then obtain
 \begin{align}
 2K_i\sin{\phi_0}\cos\Bigl(\theta_2(t-\tau_2) - \theta_1(t) + \phi_0\Bigr) \approx \omega_2 - \omega_1,  
\end{align}
which implies that $\theta_2\big(t - \tau_2\big) - \theta_1(t)$, $\theta_2\big(t - \tau_2 \big) - \theta_2(t)$,
 $\theta_1(t - \tau_1) - \theta_1(t)$, and $\theta_1(t-\tau_1) - \theta_2(t)$ are constant. Thus
\[\begin{aligned}
\dot{\theta}_1(t) 	=&\,\, \dot{\theta}_2(t) \notag \\
		=&\,\, \omega_1 - K_i\sin\Bigl(\theta_1(t-\tau_1) - \theta_1(t)\Bigr) \\
		       &\quad\,\,\, -  K_i\sin\Bigl(\theta_2(t-\tau_2) - \theta_1(t) \Bigr) \notag \\
		=&\,\, \omega_2 -   K_i\sin\Bigl( \theta_1(t-\tau_1) - \theta_2(t) \Bigr) \\
		     &\quad\,\,\, - K_i\sin\Bigl(\theta_2(t-\tau_2) - \theta_2(t)\Bigr) \notag \\
		\dot{=}&\,\,  \Omega. \notag
\end{aligned}\]
Thus from the steady state conditions
\begin{displaymath}
\begin{aligned}
 \theta_1(t)  &=\Omega{}t, \qquad \theta_2(t)  =\Omega{}t - \phi_0, \\
 \Omega &= \omega + K_i\sin\bigl(\Omega{}\tau\bigr)+K_i\sin\bigl(\Omega{}\tau+\phi_0\bigr),
\end{aligned}
\end{displaymath}
and
\[ 2K_i\sin{\phi_0}\cos\bigl(\Omega\tau\bigr) = \omega_2 - \omega_1.  \]
As $K_i$ increases, $\phi_0$ approaches zero and (\ref{eq:relIntermediateOmega2}) becomes
\[ \Omega = \omega + 2K_i\sin\bigl(\Omega{}\tau\bigr). \]

We now derive the condition for stability of the synchronised state. In~\cite{strogatzStability}, 
the synchronisation phenomena of the following time delayed coupled oscillators model is studied
\begin{align}
 \dot{\theta}_{\mym}(t) &= \omega + \frac{K}{n}\sum_{\myj=1}^{n}f\Big(\theta_\myj(t-\tau)-\theta_{\mym}(t)\Big). \label{kuramotodelay}
\end{align}
In~\cite{strogatzStability}, the in-phase ($\phi_0=0$) synchronised state  of~\eqref{kuramotodelay}, which has a collective frequency given by $\Omega=\omega + Kf(-\Omega\tau)$,  
is shown to be linearly stable if and only if
\begin{align*}
 Kf'(-\Omega\tau) > 0.
\end{align*}
Using this result, the \textit{necessary and sufficient} condition for stability  of the  synchronised state 
of the coupled oscillators~\eqref{eq:linearIntermediateEquallyCoupled} may be deduced to be 
\begin{align*}
 2K_i\cos(\Omega\tau) < 0.
\end{align*}

\end{proof}

\vspace{2ex}
\noindent\textit{Discussion:}
Note that the coupling strength $K_i$ depends on, and is proportional to, the link capacity. Thus as capacities increase, the phenomena of synchronisation will only get more pronounced.
We also note that as the value of $K_i$ increases, $\phi_0$ decreases, which results in increased phase coherence. Observe that for a large value of $K_i$, the synchronising frequency is independent of the intrinsic frequencies of the two sets of TCP flows.
\\

\subsection{Compound, Reno and Illinois}
Note that the synchronisation phenomena depends on the coupling strengths of the model.
Further, observe from (\ref{eq:couplingStrengthSmall})  and (\ref{eq:couplingStrengthInter}) that both the small and the intermediate buffer coupling strengths depend on the 
functions $i(\w{})$ and $d(\w{})$ of the transport protocols.  
We highlight the coupling strengths for  Compound, Reno and Illinois, in Table \ref{table:couplingStrengths} for both the router buffer limiting regimes. 

As Compound is widely deployed in the Internet, we aim to understand its choice of parameters in greater detail.
We consider the following cases for the Compound parameter $k$.\\

\noindent{}\textit{Case (i)}: $0<k<1$\\
In this case, the small buffer coupling strength is
$$K_s =  \frac{\alpha }{{\weq{}}^{1-k}\tau\big(1+p(\weq{})/\corep(\weq{},\weq{})\big)}.$$
Clearly, an increase in the Compound parameter $\alpha$ can result in the synchronisation of TCP flows.
Further observe that an increase in the window size $\weq{}$ would only make the  system less prone to synchronisation. 
Thus with small buffers, a higher link capacity which would in turn increase $\weq{}$, does not have a synchronising effect on the flows. \\

\noindent{}\textit{Case (ii)}: $k=0$\\
With $k=0$, we obtain
$$K_s =  \frac{\alpha}{\weq{}\tau\big(1+p(\weq{})/\corep(\weq{},\weq{})\big)}.$$
Thus an increase in the window size $\weq{}$ or the round-trip time $\tau$ would make the  system less prone to synchronisation.
Note that with $k=0$, one can get the functional forms for Reno by setting $\alpha=1$ and $\beta=0.5$ and the functional forms for Illinois by setting $\alpha=\alpha_{max}$ and $\beta=\beta_{min}$.
Hence we will have similar conditions in case of Reno and Illinois as the window size is varied.\\

\noindent{}\textit{Case (iii)}: $k=1$\\
In this case,  we have  $i(\weq{})=\alpha$, which does not depend on the equilibrium window size. Hence, we get
$$K_s =  \frac{\alpha}{\tau\big(1+p(\weq{})/\corep(\weq{},\weq{})\big)}.$$
Again, we note that increasing round-trip time $\tau$ would make the system less prone to synchronisation.
 \\
 \begin{figure}[t]
\centering
 \psfrag{100}[c]{\hspace{0.15cm}\scriptsize$100$}
 \psfrag{1000}[c]{\hspace{0.2cm}\scriptsize$1000$}
 \psfrag{10000}[c]{\hspace{0.2cm}\scriptsize$10000$}
 \psfrag{0.9990}{\hspace{-0.20cm}\scriptsize$0.9990$}
 \psfrag{0.9995}{\hspace{-0.20cm}\scriptsize$0.9995$}
 \psfrag{1.0000}{\hspace{-0.25cm}\scriptsize$1.0000$}
 \psfrag{coherence}[b][t]{{Phase coherence, $r(t)$}}
 \psfrag{capacity}[t][t]{{Link capacity, $C$ }}
 \includegraphics[height=3.2in, width=2.15in, angle=-90]{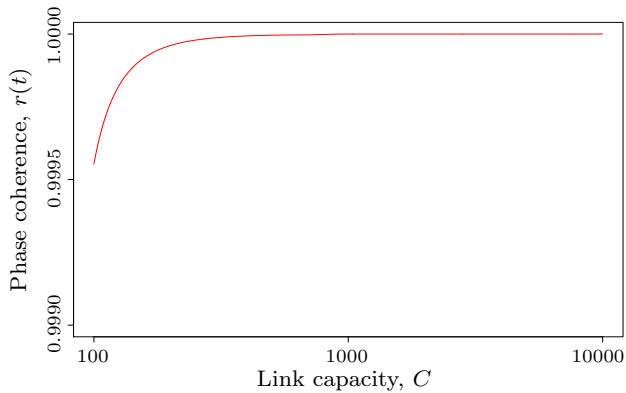} 
 \caption{\small{ Phase coherence between the oscillators in the intermediate buffer regime as the link capacities are varied. The parameters are as follows:
 the edge and core link capacities $C$ and $1.97\times C$  respectively, with round-trip times $\tau_1=0.10$  and $\tau_2=0.11$.}}
 \label{phasecohenrenceVscapacity}
\end{figure}

For intermediate buffers, Figure \ref{phasecohenrenceVscapacity} shows the variation in phase coherence, $r(t)$ for Compound TCP with respect to the change in $C$ in the intermediate buffer regime.
The coefficient of the window decrease function, $\beta$ was set to $0.5$. The core link capacity was chosen to be $1.97$ times the edge link capacity so all routers are bottlenecks. 
The round-trip times $\tau_1=0.10$ and $\tau_2=0.11$ were chosen to be close enough so that the synchronisation occurs in the core router. 

\psfrag{0}{\scriptsize$0$}
\psfrag{5}{\scriptsize$5$}
\psfrag{10}{\scriptsize\hspace{-2pt}$10$}
\psfrag{50}{\scriptsize\hspace{-2pt}$50$}
\psfrag{100}{\scriptsize\hspace{-2pt}$100$}
\psfrag{172}{\scriptsize\hspace{-2pt}$172$}
\psfrag{344}{\scriptsize\hspace{-2pt}$344$}
\psfrag{125}{\scriptsize$125$}
\psfrag{(a)}[t][b]{\hspace{1.3cm}{\small\textit{(a) Buffer size = $15$ pkts}}}
\psfrag{(b)}[t][b]{{\small\textit{(b) Buffer size = $100$ pkts}}}
\psfrag{(c)}[t][b]{\hspace{0.2cm}{\small\textit{(c) Buffer size = $344$ pkts}}}
\psfrag{reno}{\small \hspace{-0.7cm}\textit{TCP Reno}}
\psfrag{compound}[c]{\small \hspace{1.6cm}\textit{Compound TCP}}
\psfrag{illinois}[t]{\small \hspace{0.6cm}\textit{TCP Illinois}}
\psfrag{Queue}{\small \hspace{-4.5mm}\textbf{Queue size (pkts)}}

\begin{figure}[t!]
\centering
\begin{tikzpicture}
\pgfmathsetmacro{\Rsize}{20}
\pgfmathsetmacro{\Csize}{2}
\tikzset{
line width=0.6pt
}
\node[rectangle,draw=black,minimum size=\Rsize] (E1) at (1.5, 0.5*8.5) {\footnotesize $E_1$};
\foreach \i in {6,7,10,11}
{
 \node[circle,draw=black,minimum size=\Csize] (\i) at (0.5,0.5*\i) {};
 \draw[->] (\i) -- (E1);
}
\node (xs1) at (0.5, 0.5*8.5 - 0.15) {\footnotesize $\vdots$};
\node (xs2) at (0.5, 0.5*9.0 + 0.05) {\footnotesize $\vdots$};

 \node[rectangle,draw=black,minimum size=\Rsize] (E2) at (1.5,0.5*2.5-0.5) {\footnotesize $E_2$};
\foreach \i in {1,2,5,6}
{
 \node[circle,draw=black,minimum size=\Csize] (\i) at (0.5,0.5*\i-1.0) {};
 \draw[->] (\i) -- (E2);
}
\node (xs1) at (0.5, 0.5*1.5 - 0.15) {\footnotesize $\vdots$};
\node (xs2) at (0.5, 0.5*2.0 + 0.05) {\footnotesize $\vdots$};

\node (e1cap) at (2.0, 0.5*4.0) {\footnotesize $100$ Mbps};
\node (e2cap) at (2.0, 0.5*6.0) {\footnotesize $100$ Mbps};
\node (corecap) at (4.75, 0.5*5.5) {\footnotesize $197$ Mbps};
\node (crosscapsource) at (4.25, 0.5*3.5) {\footnotesize $10$ Mbps}; 
\node (crosscapsink) at   (4.25, 0.5*6.5) {\footnotesize $10$ Mbps};

\node[rectangle,draw=black,minimum size=\Rsize] (C) at (3.5, 0.5*5) {$C$};
\draw[->] (E1) -- (C);
\draw[->] (E2) -- (C);

\node[rectangle,draw=black,minimum size=\Rsize] (CT1) at (3.5, 0.5*8) {$S_{ct}$};
\draw[->] (CT1) -- (C);
\foreach \i in {1,2,5,6}
{
 \node[circle,draw=black,minimum size=\Csize] (CS\i) at (0.5*\i+1.75, 0.5*8+1.0) {};
 \draw[->] (CS\i) -- (CT1);
}
\node (y1) at (3.25, 0.5*8+1.0) {$\cdots$};
\node (y3) at (3.75, 0.5*8+1.0) {$\cdots$};

\node[rectangle,draw=black,minimum size=\Rsize] (D) at (6.0, 0.5*5) {$D$};
\draw[->] (C) -- (D);

\foreach \i in {2,3,6,7}
{
 \node[circle,draw=black,minimum size=\Csize] (d\i) at (7.0,0.5*\i+0.25) {};
 \draw[->] (D) -- (d\i);
}
\node (x1) at (7.0, 0.5*6.0 - 0.15) {$\vdots$};
\node (x3) at (7.0, 0.5*4.5 + 0.15) {$\vdots$};

\node[rectangle,draw=black,minimum size=\Rsize] (CD1) at (3.5, 0.5*2.0) {$D_{ct}$};
\draw[->] (C) -- (CD1);
\foreach \i in {2,3,6,7}
{
 \node[circle,draw=black,minimum size=\Csize] (ctds\i) at (1.25 + 0.5*\i,0) {};
 \draw[->] (CD1) -- (ctds\i);
}
\node (y1) at (3.25, 0) {$\cdots$};
\node (y3) at (3.75, 0) {$\cdots$};
\node[rotate=90] (tbulk1) at (0, 2.5) {{Bulk traffic sources}};

\node[rotate=90] (tbulk2) at (7.5, 2.5) {{Bulk traffic sinks}};

\node (tct1) at (3.5, 5.5) {{Cross traffic  sources}};

\node (tcts) at (3.5,-0.5) {{Cross traffic sinks}};

\end{tikzpicture}
\caption{ The network topology used for packet-level simulations. 
The bulk of the traffic passes through two distinct edge routers which then merge at a common core router.
The topology also consists of cross traffic comprising of both long and short-lived flows. }
\label{fig:crosstraffic}
\end{figure}
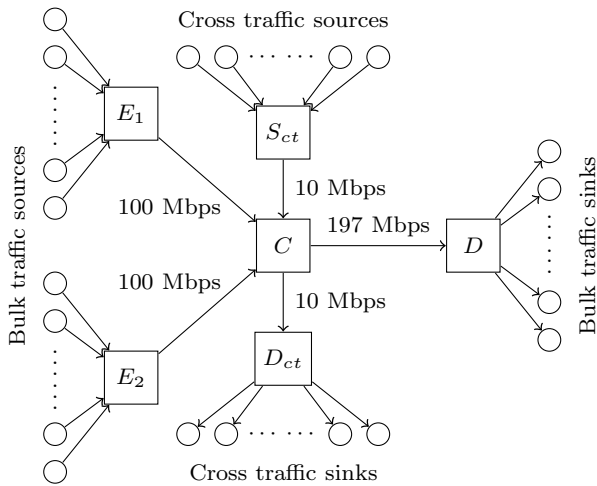

\section{Packet-level simulations}   
\label{section:simulations}
We now conduct some packet-level simulations, using version 2.35 of the Network Simulator~\cite{ns2} for Compound TCP and Illinois. The topology consists of \emph{two edge routers}, each with a capacity of $100$ Mbps, feeding into a \emph{common core router} which has capacity $197$ Mbps. Note that with this choice of capacities, we obtain a bottlenecked system. The system topology is shown in Figure~\ref{fig:crosstraffic}. 
In Figure~\ref{fig:crosstraffic}, $E_1$ and $E_2$ are the edge routers, $C$ is the core router, and $D$ is the destination for the bulk traffic sources. $S_{ct}$ and $D_{ct}$ corresponds
to the source and destination routers for the cross traffic. The router $S_{ct}$ has a capacity of $10$ Mbps.
For the bulk of the traffic, the system has two sets of $60$ long-lived TCP flows, where each flow has a $2$ Mbps access link. To incorporate burstiness in the incoming traffic, we then increase the speed of the access links, keeping the offered load fixed. To achieve this, we consider a scenario with $20$ long-lived TCP flows, each over an access link of $6$ Mbps. Each set of TCP flows goes through a separate edge router. The round-trip times of the flows are varied from $5$ ms to $200$ ms.
The difference between round-trip times of the different sets of flows were kept small.
We also introduce cross traffic which include both short-lived flows and $6$ long-lived flows, each with a $1$ Mbps access link, as shown in Figure~\ref{fig:crosstraffic}. 
For short-lived flows, it is assumed that they are all accessing a web server, and in doing so each flow generates $5$KB of data.
These short-lived HTTP flows are generated at a rate of $50$ flows per second.
The packet size is set to $1500$ bytes. In the case of small buffers, we consider two scenarios to show the impact of buffer sizes on the coupling between the two sets of flows, and in turn on the synchronization: $(i)$ Buffer sizes of all routers are fixed at $15$ packets, and (ii) Buffer sizes of all routers are fixed at $100$ packets. We plot the average window sizes of two sets of Compound TCP flows for these scenarios, see Figure 4. We can observe that, as buffer sizes increase, the system exhibits more pronounced oscillations among the TCP flows, which accurately aligns with our analytical insights. 

In the case of intermediate buffers, the prescribed dimensioning rule is $B=c*\overline{RTT}/\sqrt{N}$.
An estimate of the average round-trip time is needed to dimension router buffers. In practice, this will be highly variable but one needs to choose a fixed value to build routers. Thus, a value of $250$ ms is used for $\overline{RTT}$, by router vendors, to dimension router buffers \cite{part1}. Our analytical insights reveal that in this buffer sizing regime, increasing the capacity would increase the coupling strength, which would in turn increase the degree of synchronization among the two sets of TCP flows. 
In this buffer regime, we consider two scenarios: $(i)$ The edge routers have $270$ packets and the core router has $375$ packets, the edge capacities are fixed at $100$ Mbps, and the core router capacity is considered to be $197$ Mbps, and $(ii)$  The edge routers have $380$ packets and the core router has $530$ packets, the edge capacities are fixed at $200$ Mbps, and the core router capacity is considered to be $395$ Mbps. We now plot the average window sizes of two sets of Compound TCP flows, shown in Figure 5. We note that as capacities increase, the degree of synchronization also increases, thus validating our analysis.  

Note that this synchronization among the TCP flows leads to emergence of limit cycles in the queue size dynamics.  Figures~\ref{fig:5-10droptail},~\ref{fig:50-55droptail},~\ref{fig:100-110droptail},~\ref{fig:180-200droptail} show the queue dynamics at the core router with smooth traffic,  as the feedback delay is progressively
increased. There is a clear change in the dynamics of the queue size.
Figure \ref{fig:100-110droptail} represents the scenario where the two sets of TCP flows have round-trip times of $100$ ms and $110$ ms.
In the small buffer regime, as the buffer varies between $15$ and $100$ packets we see the onset of limit cycles in the queue size.
With intermediate buffers, we again observe non-linear oscillations. A similar insight holds even with bursty traffic, see Figure \ref{fig:100-110droptail_bursty}.
When the round-trip times were increased to $180$ and $200$ ms respectively, the limit cycles get more pronounced; see Figure \ref{fig:180-200droptail}. 
Observe from Figures \ref{fig:100-110droptail} and  \ref{fig:180-200droptail} that the intrinsic frequency depends on the transport protocol.
Also note that an increase in the round-trip time results in a decrease in the intrinsic frequency, which is in agreement with the analysis.
Note that with buffers of size $15$ packets, the system did not exhibit limit cycles in the queue~size~dynamics.

In the packet-level simulations, we observed the emergence of a synchronised state in the average window sizes among two sets of TCP flows, and also the  existence of limit cycles in the queue size dynamics, in the regime where the bandwidth-delay product is large and as the Drop-Tail buffer gets larger. An insight obtained is that in large bandwidth-delay environments, the stochastic effects of the queue get drowned out, and the non-linear dynamics of the underlying feedback control system start to dominate. 
With very small buffers, we expect a higher average loss rate. Within reason, one may prefer to deal with a higher loss rate, as compared to larger queuing delays. TCP is fairly adept at dealing with packet-losses, but queuing delays only serve to hurt everyone.


\newcommand{\myplotHeight}{3.5in}
\begin{figure*}[p]
\centering
          \psfrag{Time}[c]{\small \hspace{1cm}\begin{tabular}{c}  $\tau_1=5$ ms and $\tau_2=10$ ms \\  \textbf{Time (s)}  \end{tabular} }
          \psfrag{530}{$380$}
          \psfrag{200}{$100$}
          \psfrag{250}{$150$}
          \psfrag{40}{$40$}
            \psfrag{10}{$10$}
            \psfrag{a}{\hspace{-10mm}Buffer Size = $15$ packets}
            \psfrag{aaa}{Set $1$}
             \psfrag{xyz}{Set $2$}
             \psfrag{b}{\hspace{-20mm}Buffer Size = $100$ packets}
          \psfrag{y}{\hspace*{2mm}Average Window Size (pkts)}
          \psfrag{x}{\hspace{-5mm}Time (seconds)}
          \psfrag{illinois}{\vspace{-2mm}\hspace{-6mm}\textit{TCP Illinois}}
                \includegraphics[height=15cm, width=7.5cm, angle=-90]{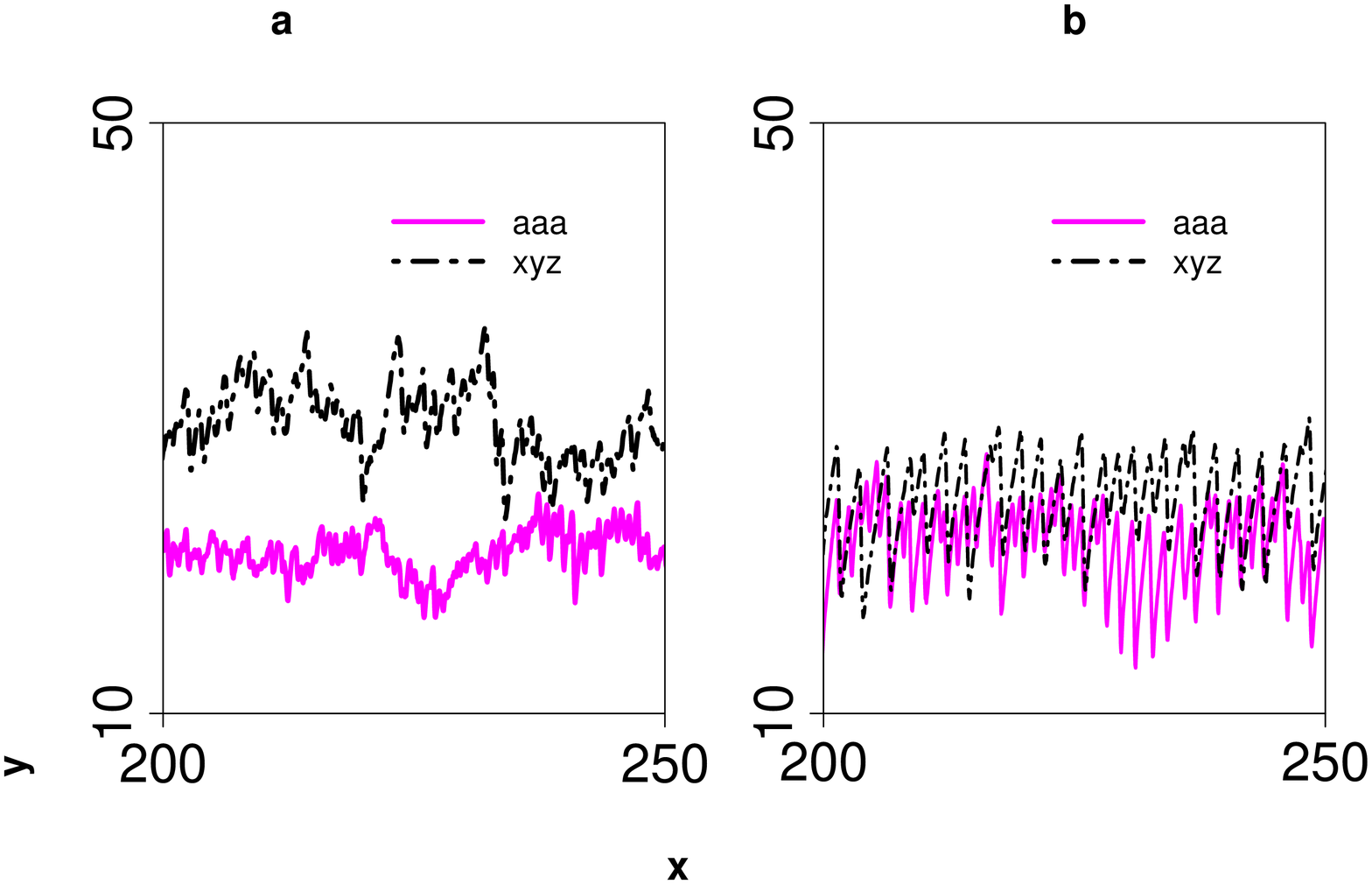} 
                \caption{Packet-level simulation traces of average window sizes with smooth traffic, for small buffer regime. Observe that for larger buffers, the degree of synchronization is higher. }
                \label{fig:15-100droptail}
\end{figure*}

\begin{figure*}[p]
\centering
          \psfrag{Time}[c]{\small \hspace{1cm}\begin{tabular}{c}  $\tau_1=5$ ms and $\tau_2=10$ ms \\  \textbf{Time (s)}  \end{tabular} }
          \psfrag{530}{$530$}
          \psfrag{200}{$100$}
          \psfrag{250}{$150$}
          \psfrag{40}{$40$}
            \psfrag{10}{$10$}
            \psfrag{a}{\hspace{-10mm}Edge capacity = $100$ Mbps}
            \psfrag{aaa}{Set $1$}
             \psfrag{xyz}{Set $2$}
             \psfrag{b}{\hspace{-20mm}Edge capacity = $200$ Mbps}
          \psfrag{y}{\hspace*{2mm}Average Window Size (pkts)}
          \psfrag{x}{\hspace{-5mm}Time (seconds)}
          \psfrag{illinois}{\vspace{-2mm}\hspace{-6mm}\textit{TCP Illinois}}
                \includegraphics[height=15cm, width=7.5cm, angle=-90]{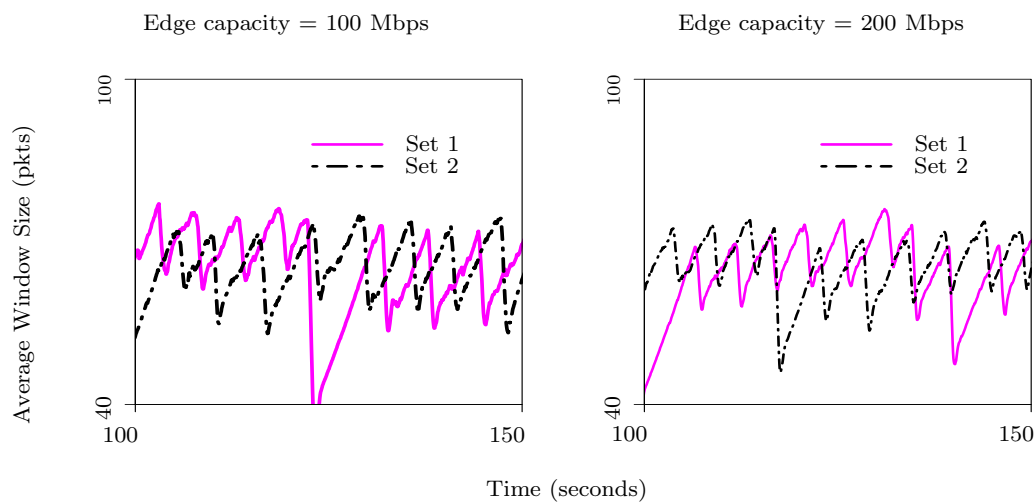} 
                \caption{Packet-level simulation traces of average window sizes with smooth traffic, for intermediate buffer regime. Observe that for larger capacities, the degree of synchronization is higher.  }
                \label{fig:100-200droptail}
\end{figure*}
\begin{figure*}[p]
\centering
          \psfrag{Time}[c]{\small \hspace{1cm}\begin{tabular}{c}  $\tau_1=5$ ms and $\tau_2=10$ ms \\  \textbf{Time (s)}  \end{tabular} }
          \psfrag{530}{$375$}
          \psfrag{100}{$100$}
          \psfrag{125}{$125$}
          \psfrag{0}{$0$}
            \psfrag{10}{$15$}
            \psfrag{(c)}{\hspace{-16mm}\small\textit{\vspace{1cm}(c) Buffer size = $375$ pkts }}
          \psfrag{Queue}{\small\hspace{0.75cm}\textbf{Queue Size (pkts)}}
          \psfrag{illinois}{\vspace{-2mm}\hspace{-6mm}\textit{TCP Illinois}}
                \includegraphics[height=15cm, width=15cm, angle=-90]{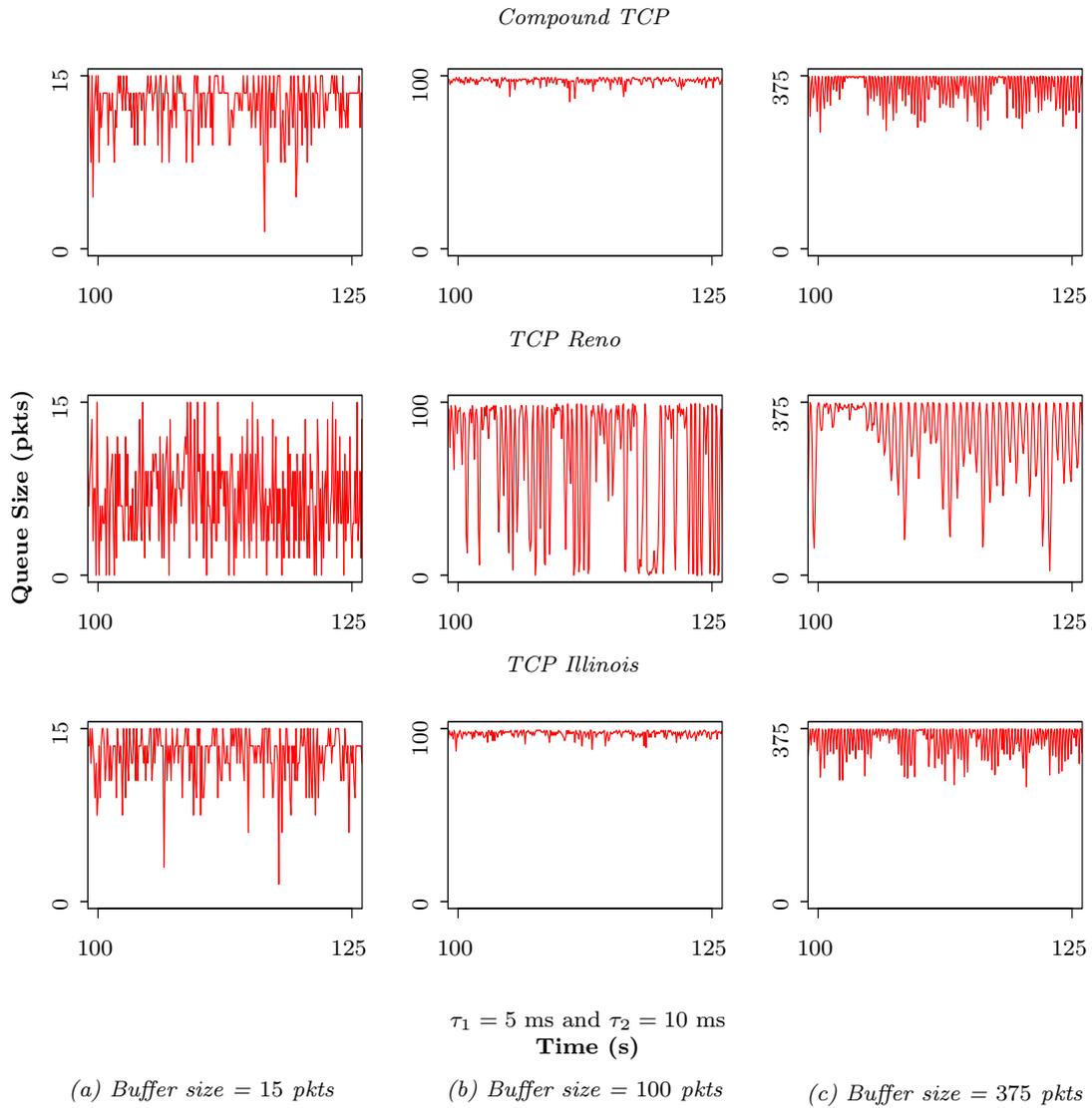} 
                \caption{Packet-level simulation traces at the \textit{core} router with smooth traffic, for average round-trip times $\tau_1=5$ ms and $\tau_2=10$ ms. Observe that for such small round-trip times, the queues do not exhibit oscillations.  }
                \label{fig:5-10droptail}
\end{figure*}
\begin{figure*}[p]
\centering
          \psfrag{Time}[c]{\small \hspace{1cm}\begin{tabular}{c}  $\tau_1=50$ ms and $\tau_2=55$ ms \\  \textbf{Time (s)}  \end{tabular} }
          \psfrag{530}{$375$}
          \psfrag{100}{$100$}
          \psfrag{125}{$125$}
          \psfrag{0}{$0$}
            \psfrag{10}{$15$}
            \psfrag{(c)}{\hspace{-16mm}\small\textit{\vspace{1cm}(c) Buffer size = $375$ pkts }}
          \psfrag{Queue}{\small\hspace{0.75cm}\textbf{Queue Size (pkts)}}
          \psfrag{illinois}{\vspace{-2mm}\hspace{-6mm}\textit{TCP Illinois}}
                \includegraphics[height=15cm, width=15cm, angle=-90]{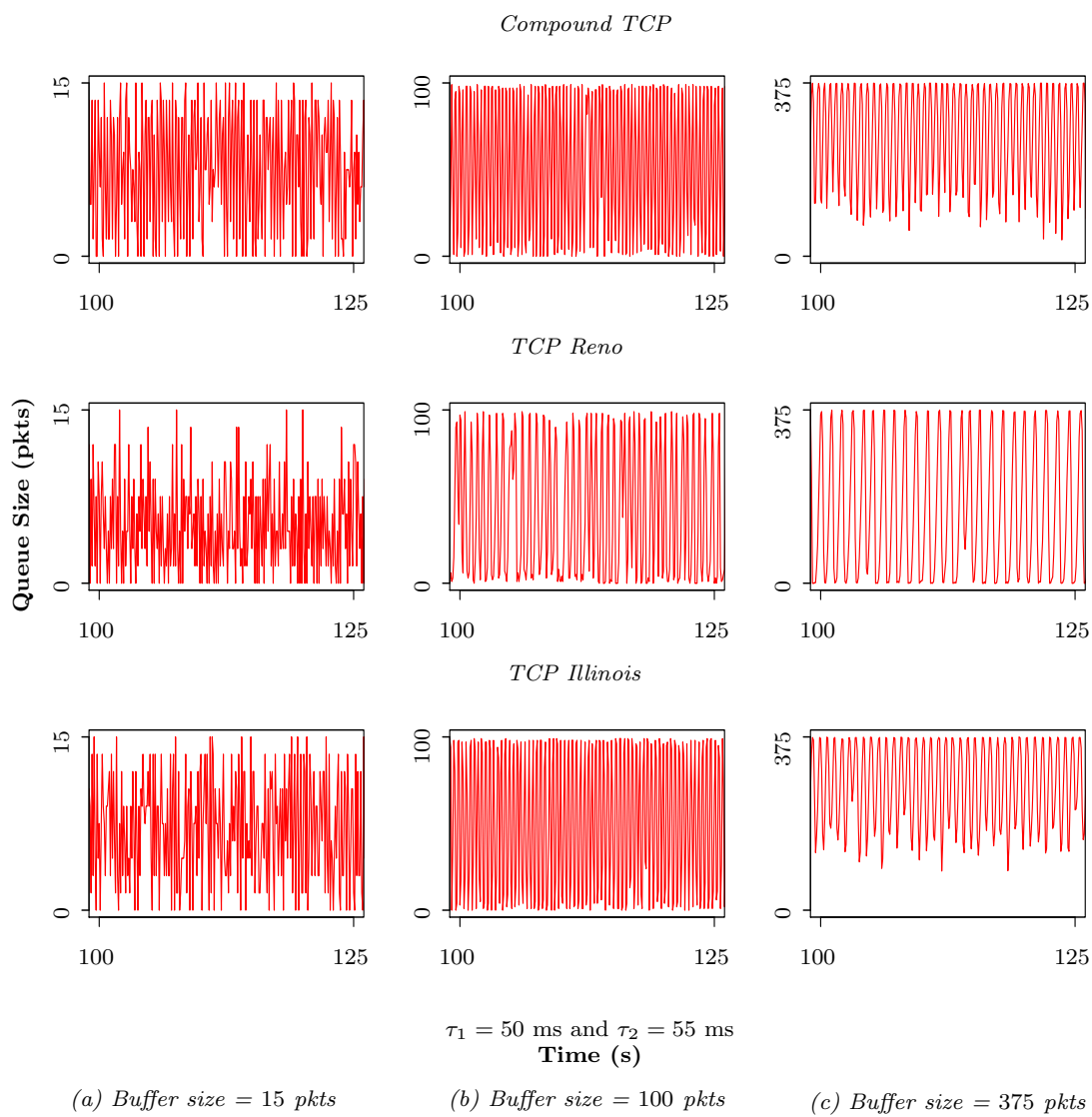}  \caption{ Packet-level simulation traces at the \textit{core} router with smooth traffic, for average
                round-trip times $\tau_1=50$ ms and $\tau_2=55$ ms. Observe the onset of oscillations in the queue size dynamics as the buffer size is increased. }
                \label{fig:50-55droptail}
\end{figure*}
\begin{figure*}[p]
\centering
         \psfrag{Time}[c]{\small \hspace{1cm}\begin{tabular}{c}  $\tau_1=100$ ms and $\tau_2=110$ ms \\  \textbf{Time (s)}  \end{tabular} }
          \psfrag{530}{$375$}
          \psfrag{100}{$100$}
          \psfrag{125}{$125$}
          \psfrag{0}{$0$}
            \psfrag{10}{$15$}
            \psfrag{(c)}{\hspace{-16mm}\small\textit{\vspace{1cm}(c) Buffer size = $375$ pkts }}
          \psfrag{Queue}{\small\hspace{0.75cm}\textbf{Queue Size (pkts)}}
          \psfrag{illinois}{\vspace{-2mm}\hspace{-6mm}\textit{TCP Illinois}}
                \includegraphics[height=15cm, width=15cm, angle=-90]{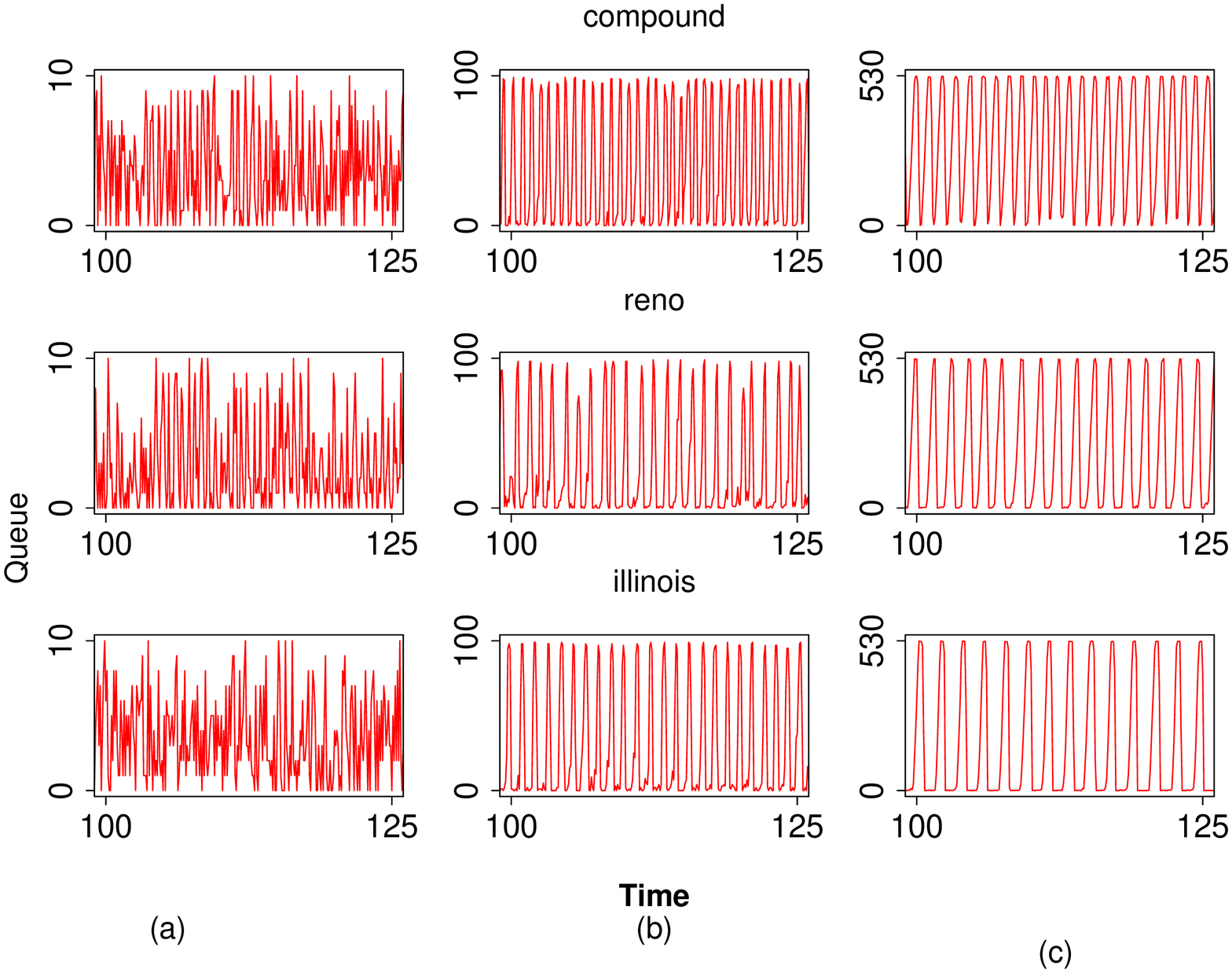}  \caption{  Packet-level simulation traces at the \textit{core} router with smooth traffic, for average
                round-trip times $\tau_1=100$ ms and $\tau_2=110$ ms. As expected from the analysis, note the emergence of limit cycles with a buffer size of $100$ packets, and in the intermediate buffer regime ($375$ packets). 
                 As expected from the analysis, the intermediate buffer regime ($375$ packets) also has non-linear oscillations. }
                \label{fig:100-110droptail}
\end{figure*}
\begin{figure*}[p]
\centering
          \psfrag{Time}[c]{\small \hspace{1cm}\begin{tabular}{c} {\small $\tau_1=180$ ms and $\tau_2=200$ ms} \\  \textbf{Time (s)}  \end{tabular} }
          \psfrag{530}{$375$}
          \psfrag{100}{$100$}
          \psfrag{125}{$125$}
          \psfrag{0}{$0$}
            \psfrag{10}{$15$}
            \psfrag{(c)}{\hspace{-16mm}\small\textit{\vspace{1cm}(c) Buffer size = $375$ pkts }}
          \psfrag{Queue}{\small\hspace{0.75cm}\textbf{Queue Size (pkts)}}
          \psfrag{illinois}{\vspace{-2mm}\hspace{-6mm}\textit{TCP Illinois}}
                \includegraphics[height=15cm, width=15cm, angle=-90]{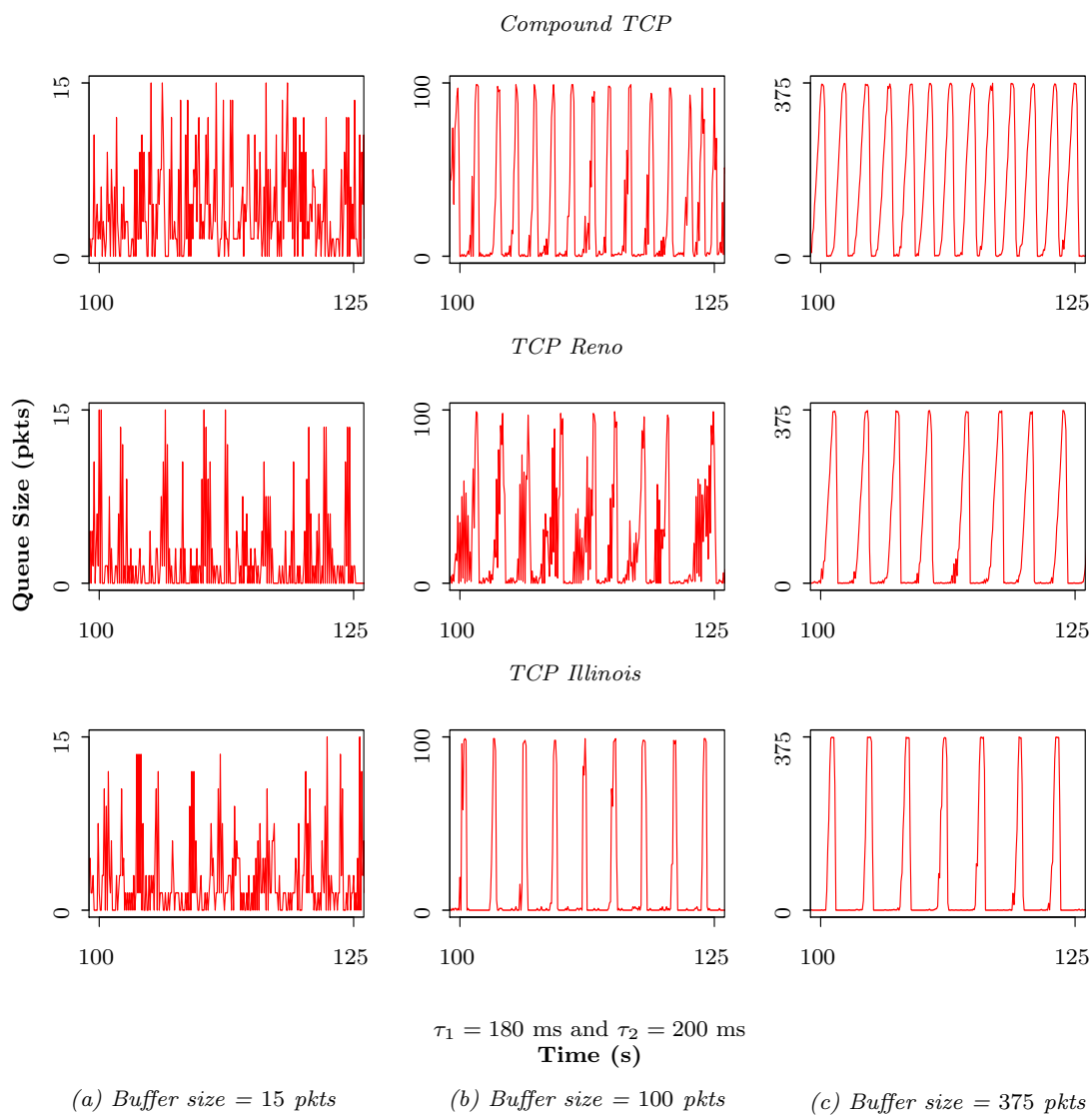} \caption{  Packet-level simulation traces at the \textit{core} router with smooth traffic, for average
                round-trip times $\tau_1=180$ ms and $\tau_2=200$ ms. As compared to Figure~\ref{fig:100-110droptail}, the non-linear oscillations are more pronounced as the feedback time delays
                 are larger.   }
                \label{fig:180-200droptail}
\end{figure*}
\begin{figure*}[p]
\centering
         \psfrag{Time}[c]{\small \hspace{1cm}\begin{tabular}{c}  $\tau_1=100$ ms and $\tau_2=110$ ms \\  \textbf{Time (s)}  \end{tabular} }
          \psfrag{530}{$375$}
          \psfrag{100}{$100$}
          \psfrag{125}{$125$}
          \psfrag{0}{$0$}
            \psfrag{10}{$15$}
            \psfrag{(c)}{\hspace{-16mm}\small\textit{\vspace{1cm}(c) Buffer size = $375$ pkts }}
          \psfrag{Queue}{\small\hspace{0.75cm}\textbf{Queue Size (pkts)}}
          \psfrag{illinois}{\vspace{-2mm}\hspace{-6mm}\textit{TCP Illinois}}
                \includegraphics[height=15cm, width=15cm, angle=-90]{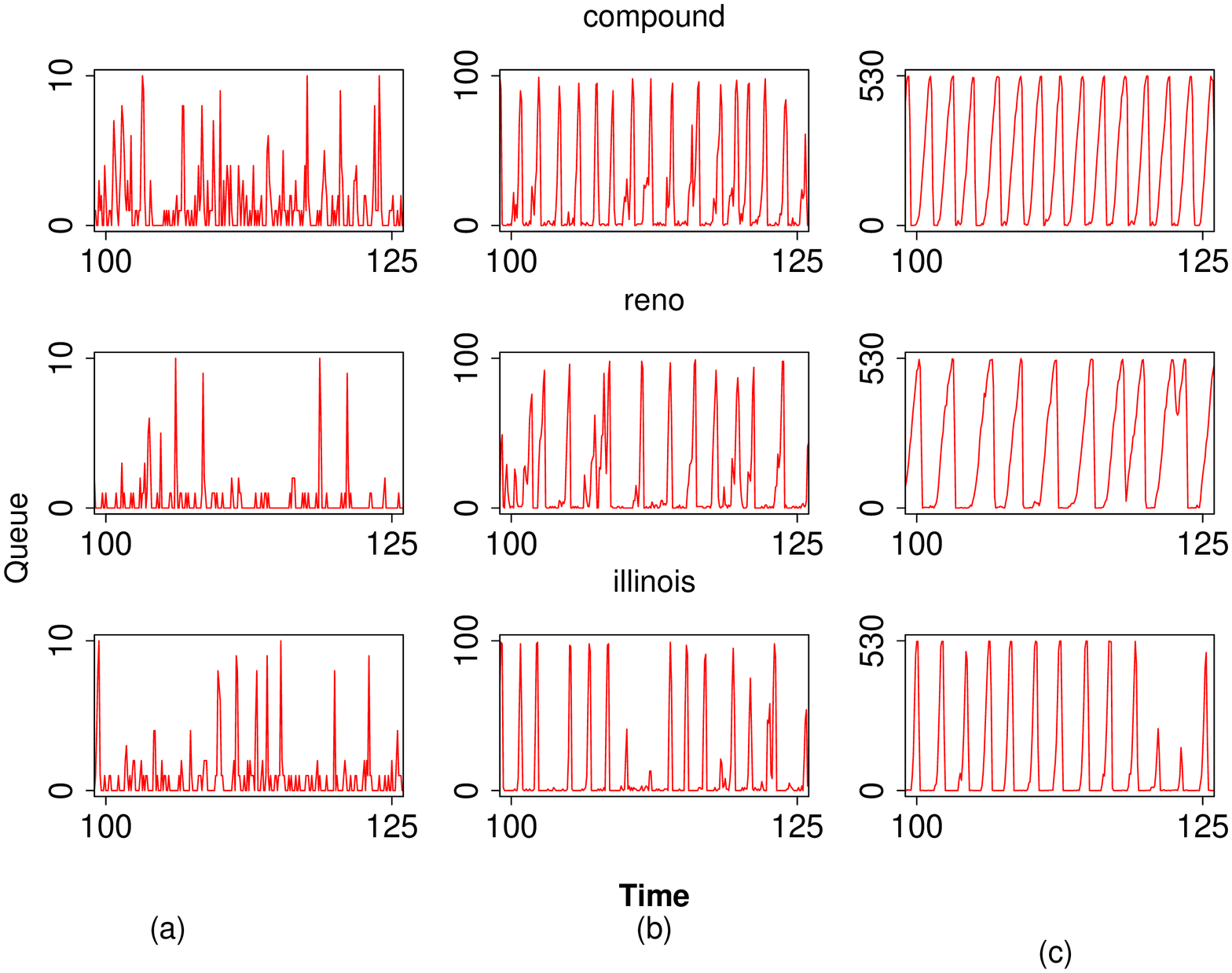}  \caption{  Packet-level simulation traces at the \textit{core} router with bursty traffic, for average round-trip times $\tau_1=100$ ms and $\tau_2=110$ ms. As expected from the analysis, note the emergence of limit cycles with a buffer size of $100$ packets, and in the intermediate buffer regime ($375$ packets). 
                 As expected from the analysis, the intermediate buffer regime ($375$ packets) also has non-linear oscillations. }
                \label{fig:100-110droptail_bursty}
\end{figure*}
\section{Contributions}
\label{section:conclusion}

We analysed fluid models of Compound, Reno and Illinois coupled with Drop-Tail, in a small and an intermediate buffer regime. As Compound TCP and Drop-Tail queues are both widely implemented, they form an important part of network performance evaluation.
%
%
The style of analysis we employ, for the fluid models, is inspired by time delayed coupled oscillators~\cite{strogatz}. A key focus was on \emph{synchronisation} between the two sets of TCP flows, where each set had a different feedback delay. The two sets of flows were regulated by separate edge routers, which then merge at a common core router. For both buffer sizing regimes the condition under which synchronisation occurs were made explicit in terms of \emph{coupling strengths}.                                    

With \emph{small buffers}, the coupling strength depends on both protocol and network parameters. In particular, larger buffer sizes readily induce synchronisation. With \emph{intermediate buffers}, the coupling strength is directly proportional to the link capacity. So high-speed networks, with intermediate buffers and Drop-Tail would be prone to synchronisation. 
We conducted packet-level simulations to validate some of our analytical insights. Synchronisation, in the form of limit cycles, was indeed observed in the queue size with both larger buffer sizes and feedback delays. In high-speed networks, small buffers can be chosen to ensure low-latency and system stability. So in terms of network design, small buffers appear more appealing than intermediate buffers; at least with Drop-Tail queue management.             

Going forward, it would be valuable to study the underlying system when different queue policies, like 
the well-known RED~\cite{red} and REM~\cite{rem} policies, are employed at the bottleneck queues. Such studies would naturally lead to a better and quantitative understanding of network performance and dynamics.

\end{document}